\documentclass[useAMS,usenatbib]{mn2e}

\usepackage{graphicx,amssymb}
\usepackage[normalem]{ulem}
\usepackage[usenames,dvipsnames]{xcolor}
\usepackage{soul}

\title[SN blasts propelling WD-polluting debris]
{Resupplying planetary debris to old white dwarfs with supernova blast waves}

\author[]{Dimitri Veras$^{1,2,3}$\thanks{E-mail: dimitri.veras@aya.yale.edu}
\\
$^{1}$Centre for Exoplanets and Habitability, University of Warwick, Coventry CV4 7AL, UK
\\
$^{2}$Centre for Space Domain Awareness, University of Warwick, Coventry CV4 7AL, UK
\\
$^{3}$Department of Physics, University of Warwick, Coventry CV4 7AL, UK
}

\begin{document}
\label{firstpage}
\pagerange{\pageref{firstpage}--\pageref{lastpage}}
\maketitle

\begin{abstract}
One challenge with explaining how high levels of planetary debris can enrich, or ``pollute", old ($\sim$3 Gyr) and very old ($\sim$10 Gyr) white dwarfs is that debris reservoirs deplete on shorter timescales, akin to the solar system's already eviscerated Main Belt and Kuiper Belt. Here, I explore how these extrasolar reservoirs can be resupplied through supernovae that propel distant ($\gtrsim 10^4$~au) dust, sand and pebbles, and potentially boulders and comets, into the inner ($\lesssim 10^2$~au) planetary system. I analytically constrain the geometry of these blast waves, and derive expressions for the probability of apt blast configurations occurring. I then derive the minimum kick magnitudes needed to generate stable, leaky and broken post-blast orbits, and prove that within this formalism, at most 23 per cent of true anomalies along an eccentric orbit could allow for resupplied planetary debris to experience repeated pericentre passages.  By linking these kick magnitudes with debris sizes and relating these quantities to the local neighbourhood supernova rate, I conclude that the probabilities for ejection or resupply per supernova blast are $\approx$100 per cent for micron-sized dust and millimetre-sized pebbles and sand, and $\approx$0 per cent for asteroids larger than $\sim$10~km. In between these extremes, I expect metre-sized boulders to be resupplied at least once to very old white dwarfs over their cooling ages. The efficacy of this debris delivery mechanism is dependent on the time-varying sources and sinks in an exo-Oort cloud and how its parent white dwarf has, throughout its cooling age, traversed the Milky Way.
\end{abstract}


\section{Introduction}

As of June 2025, over a quarter of all $\gtrsim$ 6,200 known planetary systems host white dwarfs that contain observable planetary remnants in their atmospheres\footnote{The systems with known exoplanets, which include over 4,400 systems, are catalogued in the NASA Exoplanet Archive, at https://exoplanetarchive.ipac.caltech.edu/, and the Encyclopaedia of Exoplanetary Systems, at https://exoplanet.eu/home/.} \citep{wiletal2024}. These remnants, in the form of dissociated debris, are said to pollute or enrich the white dwarf, and notably reveal the chemical composition of its progenitor through spectroscopic analysis \citep[e.g.][]{bonsor2024,xuetal2024}.

Enriched white dwarfs vary in age by several orders of magnitude, with the oldest having already existed as a white dwarf for $\sim 10$~Gyr \citep{elmetal2022}. Although the actual functional dependence of accretion rate of planetary debris versus white dwarf cooling age remains uncertain \citep{holetal2018,cheetal2019,bloxu2022,manetal2024}, what is certain is that planetary debris needs to exist in sufficiently accessible reservoirs around white dwarfs over $\sim 10$~Gyr timescales in order to explain the observations.

However, in the solar system, two main sources of debris that could enrich the eventual solar white dwarf are the Main Belt and the Kuiper Belt \citep{lietal2022}. Both of these repositories have already been severely depleted \citep{gladman1997,morbidelli1997,botetal2005,tismal2009,graetal2017,botetal2023}, despite the Sun having over 6 Gyr remaining before leaving the main sequence.

The subsequent new supply of debris which may be created due to destructive giant branch radiation \citep{veretal2014c,versch2020} will similarly start to thin out as the solar white dwarf ages. Whatever debris that remains will be gravitationally influenced by surviving planets and white dwarf radiation.  The solar system's surviving planets (at least Mars, Jupiter, Saturn, Uranus, Neptune) will approximately double their semi-major axes, while largely otherwise keep their existing orbital shape \citep{veretal2011}. In extrasolar systems with higher main-sequence host star masses, the surviving planets could instead triple their semi-major axes during the star's transformation into a white dwarf.

Exoplanets which survive giant branch evolution and end up at distances within $\sim$100~au of their parent white dwarf have long been invoked as the primary gravitational driver of debris into the white dwarf (\citealt*{bonetal2011}, \citealt*{debetal2012}, \citealt*{frehan2014}, \citealt*{musetal2018}, \citealt*{smaetal2018}, \citealt*{broetal2022}, \citealt*{ocoetal2022}, \citealt*{verros2023}, \citealt*{lietal2025}; see \citealt*{veretal2024} for a more complete listing of investigations). Radiation from young and hot white dwarfs alone could also work in the same $\sim$100~{\rm au} region to drag material into the white dwarf \citep{veretal2022}, although as the white dwarf cools, this region shrinks.  However, the planets will remain. Hence, over long cooling timescales of the white dwarf, planetary debris needs to be supplied to this $\sim$100~au ``perturbation zone". Here, the existing architecture can complete the debris delivery to the white dwarf \citep{bonwya2012,wyaetal2017,maretal2018,rodlai2024} or at least propel the debris close enough for white dwarf radiation to eventually complete the delivery through Poynting-Robertson drag. 

A potential source for resupplying the perturbation zone is an exo-Oort cloud, which is likely to be $\sim 10^3-10^5$~au distant from their parent stars. These clouds are formed by both gravitational scattering events subsequent to the dissipation of the nascent protoplanetary disc, as well as with interactions from other stars in the stellar birth cloud \citep{dunetal1987,donetal2015,higkok2015,baifab2019,corcal2019,poretal2021,rayetal2023,wajetal2024}. These distant reservoirs are continuously shifting, and are being depleted and repleneshed throughout the star's evolution \citep{belraf2010,vereva2013,saillenfest2020,penarrubia2023}; in fact, we might expect $\sim 10^{14}$ interstellar visitors like 1I/‘Oumuamua or 2I/Borisov to be passing through the solar system's $10^5$~au-scale Oort cloud at any instant \citep{jewitt2017,trietal2017,doetal2018,jewsel2023}.

Can exo-Oort cloud objects orbiting white dwarfs be thrust into the perturbation zone? This question has been investigated for over four decades, and researchers have invoked natal white dwarf kicks, Galactic tides, stellar flybys, passing molecular clouds, and Alfv\'{e}n-wave drag to demonstrate that this transfer is possible \citep{alcetal1986,paralc1998,veretal2014b,stoetal2015,caihey2017,zhaetal2021,ocoetal2023,akietal2024,pharei2024}. However, Galactic supernovae have scarcely been mentioned as a potentially important driver\footnote{See \cite{steshu1988} for a treatment of how supernovae thermally affect exo-Oort clouds; potential sublimation of comets could represent fresh sources of debris throughout an old star's lifetime \citep{belpav2024}.}, and hence is the subject of this work.

Supernovae could only be important in this context if they occur both frequently enough and close enough to passing white dwarfs that harbour exo-Oort clouds. A recent census of core-collapse supernova rates in the Solar neighbourhood \citep{quietal2025} reveals that a supernova occurs at a distance of $\sim 20$~pc from the Sun every $\sim$2.5~Gyr. Further, \cite{firestone2014} suggested that nearly two dozen supernovae occurred within 300~pc of the Sun within just the last 300~Myr. In fact, recent, nearby and multiple supernovae have been linked to the geological record on Earth \citep{benetal2002,breetal2016,waletal2016,fieetal2020,milfie2022}.

These rates and proximities do suggest that supernovae could be a prevalent driver of white dwarf enrichment for old white dwarfs by thrusting distant material into the perturbation zone. Supporting this notion is the study by \cite{smietal2024}, who recently demonstrated that supernovae can perturb Oort cloud material into the inner solar system. 

The impulsive nature of supernovae, combined with the distances involved, facilitate the use of analytics for investigation. Here, I use the impulse formulation established by \cite{jacetal2014} as the foundation for my derivations. In Sections 2 and 3, I detail the preliminaries by justifying my variable and parameter choices for the investigated planetary systems and supernova blast waves. I then step through the derivations in Sections 4-7 and support these results with numerical tests in Section 8. In Section 9 I make the link with the local neighbourhood supernova rate, and then I finally discuss the results in Section 10, and conclude in Section 11.

\section{Orbital elements and physical parameters}

In this and the subsequent section I outline the relevant variables and parameters for the planetary systems studied. Because the analytic treatment is general, specific numerical values provided here will not be used until Section 8.

Planetary objects and debris that are emplaced into, or are just passing through, distant regions of stellar systems vary in size and type. This material includes dust as well as pebbles, cobbles, boulders, comets and even some objects perhaps better characterised as asteroids \citep{weilev1997,shaetal2015}. Given this variety, I choose to refer to all such bodies as ``small objects" or ``small bodies" for the remainder of the paper. 

The small bodies are assumed to be orbiting a white dwarf with a typical mass of $M_{\star} = 0.6M_{\odot}$. This stellar mass value remains fixed throughout the paper, and this choice will only weakly affect the results. Furthermore, the mass distribution of white dwarfs is sharply peaked around $0.6M_{\odot}$ such that slightly changing the value to, e.g., $0.4M_{\odot}$ or $0.8M_{\odot}$ would clearly not be representative of most single white dwarfs \citep[e.g.][]{cunetal2024}.

The orbital parameters most relevant for this study are the small body's semi-major axis $a$, eccentricity $e$, orbital pericentre $q$, orbital apocentre $Q$ and true anomaly $f$. The instantaneous separation between the white dwarf and small body $r$, is given by 

\begin{equation}
r = \frac{a \left(1 - e^2\right)}{1 + e \cos{f}}
\label{dist}
\end{equation}

\noindent{}such that $r=q$ when $f=0$ and $r=Q$ when $f=\pi$. Although an analysis of the evolution of the small body's inclination, argument of pericentre and longitude of ascending node may also be performed, that would represent a digression; the topic of interest here is contorting small body orbits into the perturbation zone, which largely involves just $a$, $e$, $q$ and $f$. 

For these variables, I consider two temporal states: pre-blast and post-blast. The pre-blast state, or the initial state, is earmarked with subscripts of ``i", and the post-blast state, or the final state, is indicated by subscripts of ``f".

The applicable range of $r_{\rm i}$ is beyond the location where the impulse approximation can be applied but within the Hill ellipsoid of the stellar system, such that 

\begin{equation}
r_{\rm imp} \le r_{\rm i} \le r_{\rm esc}. 
\end{equation}

\noindent{}The impulse approximation is viable when the orbital period of the object is much larger than the duration of the impulse. \cite{smietal2024} estimate the latter to be $\sim 0.05$~{\rm Myr}. This timescale, when equated to a single orbit around a $0.60M_{\odot}$ white dwarf, corresponds to a semi-major axis of $\approx $ 1,140~au. An orbital period which is 20 times longer increases this semi-major axis value to $\approx $ 8,430~au. Hence, I adopt $r_{\rm imp} = $ 9,000~au. Note importantly that although I require $r_{\rm i} \ge r_{\rm imp}$, the value of $q_{\rm i}$ may be much less than $r_{\rm imp}$ at the time of the supernova.

In order to determine $r_{\rm esc}$, consider the dimensions of the typical Hill ellipsoid of a planetary system. The dimensions of this ellipsoid are set largely by the distance to the Galactic centre, and will vary as a star migrates throughout the Galaxy. For a Solar analogue which is 8~kpc away from the Galactic centre, the shortest dimension of this ellipsoid is about 149,000~au \citep{vereva2013,veretal2014a}. A $0.6M_{\odot}$ star reduces this value by $1-0.6^{(1/3)}=$ 16 per cent. Further, most observed white dwarfs are (currently) in the Solar neighbourhood, well within 1~kpc of the Sun \citep{rebetal2019,kiletal2020,genetal2021,toretal2021,obretal2024,robetal2025}. Hence, I adopt $r_{\rm esc} = $ 125,000~au. Note that at the time when the supernova blast wave reaches the small body, the osculating value of $Q_{\rm i}$ may be much larger than $r_{\rm esc}$, although it must be the case that $r_{\rm i}\le r_{\rm esc}$.

The other key distance to consider is the extent of the perturbation zone, which I define as $r_{\rm pert} = 100$~au. This value is arguably conservatively low. Exo-debris discs around main-sequence stars can extend to distances of a few hundred au \citep{marino2022,pearce2024}, and massive but unusual planets like HR 8799~b are currently located at $\sim 70$~au \citep{maretal2010}; if that planet survives the giant branch phases, it will end up at a distance of $\approx 200$~{\rm au} around the resulting white dwarf, with both remnant debris and potentially additional planets in tow \citep{verhin2021}.

Because the small bodies significantly vary in size, I adopt a wide range of radii $R$, from $10^{-6}$~m to $10^4$~m. This range corresponds to micron-sized dust on one end and small 10~km-scale comets/asteroids on the other. This lower bound is feasible because micron-sized grains are not subject to either white dwarf winds (as white dwarfs have no winds) or radiative blow-out, given the dimness of the star. To demonstrate that micron-sized dust is not susceptible to blow-out, I write the blow-out size for a particle, $R_{\rm blow}$, as \citep{buretal1979,broetal2022}

\[
R_{\rm blow} = 1.5 {\mu}m \left( \frac{a}{\rm au} \right) \left( \frac{r}{R_{\odot}} \right)^{-1}
\left( \frac{L_{\star}}{0.01 L_{\odot}} \right)
\]

\[
\ \ \ \ \ \ \ \ \ \ \ \ \ \ \times
\left( \frac{M_{\star}}{0.6 M_{\odot}} \right)^{-1}
\left( \frac{\rho}{2.7 {\rm g} \, {\rm cm}^3} \right)^{-1}
\]

\[
\ \ \ \ \ \ \ \ \
\sim 0.15 {\mu}m \left( \frac{R_{\odot}}{\rm au} \right)
                 \left( \frac{L_{\star}}{0.001L_{\odot}} \right)
\]
                 
\begin{equation}                 
\ \ \ \ \ \ \ \ \
\lesssim 10^{-3} {\mu}m,
\label{blowout}
\end{equation}

\noindent{}where $L_{\star}$ is the luminosity of the white dwarf and $\rho$ is the density of the small object. Hence, micron-sized dust is not radiatively blown out of old white dwarf planetary systems. 

The value of the small body density will also need to be used later. I henceforth adopt $\rho = 1.5~{\rm g} \, {\rm cm}^{-3}$. This value is representative of the rocks on the near-Earth Apollo group asteroid Ryugu \citep{heretal2021}; this choice will only weakly affect the results, especially given the wide variation in small body radii that I consider.

\section{The supernova blast wave}

I consider a supernova which occurs at a distance of $D_{\rm SN}$ from the white dwarf and with an energy of $E_{\rm SN} = 10^{44}$~J. This energy value is a standard approximation used throughout the literature \citep[e.g.][]{ruisei2025}, and remains fixed for this paper. I allow $D_{\rm SN}$ to vary from pc-scale values to kpc-scale values.

The supernova blast wave will produce a ``kick" that is assumed to have a magnitude of $\Delta v$ and is propelled in a direction given by the polar angle $\phi$ and vertical angle $\theta$. Without loss of generality, I orient the pre-blast orbital plane of the small body to be the $x$-$y$ plane with the $x$-axis pointing in the direction of the orbital pericentre, as is standard convention. Then, the angle $\phi$ is the angle between the $x$-axis and the projection of the blast direction into the $x$-$y$ plane, while $\theta$ is the angle between the blast direction and the $z$-axis (see Fig. 1 of \citealt*{jacetal2014}).

The following values of $\theta$ and $\phi$ then define different types of specific kicks which are useful to characterise as the following:

\begin{itemize}

\item {\it Normal Kick}. When $\theta = 0$ or $\theta = \pi$, the kick is normal to the orbit of the small body.

\

\item {\it Coplanar Kick}. When $\theta = \pi/2$, the kick is co-planar to the orbit of the small body.

\

\item {\it Radial Kick}. When $\phi = f_{\rm i}$, the kick is in the radial direction towards the small body within the body's orbital plane.

\

\item {\it Anti-radial Kick}. When $\phi = f_{\rm i} + \pi$, the kick is in the radial direction away from the small body within the body's orbital plane.

\

\item {\it Tangential Kick}. When $\phi = f_{\rm i} + \pi/2$, the kick is tangent to the orbit of the small body, in its orbital plane, and in the direction of motion.

\

\item {\it Anti-tangential Kick}. When $\phi = f_{\rm i} - \pi/2$, the kick is tangent to the orbit of the small body, in its orbital plane, and in the opposite direction of motion.

\end{itemize}

The value of $\Delta v$ incorporates two of the most important quantities, $D_{\rm SN}$ and $R$. This $\Delta v$ value can be obtained by equating some fraction, $\gamma$, of the energy imparted by the supernova blast wave to the resulting kinetic energy of the small body. This fraction is dependent on (i) how much of the supernova energy incident upon the small body is absorbed, and (ii) how much of the absorbed energy becomes kinetic energy. Hence,

\begin{equation}
\gamma \left( \frac{E_{\rm SN}}{4\pi D_{\rm SN}^2} \right) \left(2 \pi R^2 \right)
=
\frac{1}{2} m \left(\Delta v\right)^2
\approx
\frac{2}{3} \pi R^3 \rho \left(\Delta v\right)^2,
\end{equation}

\noindent{}where $m$ is the mass of the small body. Therefore,

\begin{equation}
\Delta v = \sqrt{\frac{3 \gamma E_{\rm SN}}{4\pi \rho R D_{\rm SN}^2}}.
\label{blastwave}
\end{equation}

\noindent This expression is equivalent to that of \cite{smietal2024} when $\gamma = 1/2$. Here, given the lack of a specific physical justification to suggest otherwise, I adopt $\gamma=1$. 

Note that the supernova supplies the energy, and how that energy is translated into the kick magnitude depends on a number of variables. In this sense, I treat all supernova explosions here as equivalent (all with $E_{\rm SN} = 10^{44}$~J). 

A related quantity which will appear repeatedly in the analytics is the ratio $\Delta v/v_{\rm c,i}$, where the initial circular velocity of the small body, $v_{\rm c,i}$, is

\begin{equation}
v_{\rm c,i} = \sqrt{\frac{G M_{\star}}{a_{\rm i}}}.
\end{equation}

%\noindent{}My strategy will be to constrain this ratio $\Delta v/v_{\rm c,i}$. Then, after determining its allowable ranges, these ranges can be related to the quantities $D_{\rm SN}$ and $R$.

\section{How kicks shrink and enlarge orbits}

To begin the analytical treatment, I first consider how a supernova kick changes the semi-major axis of the small body. The post-blast semi-major axis of the small body, $a_{\rm f}$, can be expressed entirely in terms of the body's initial orbital elements through \citep{jacetal2014}

\[
a_{\rm f} = a_{\rm i} 
\bigg[1 - \left( \frac{\Delta v}{v_{\rm c,i}} \right)^2  
\]      
     
\begin{equation} 
\ \ \ \ \ \ \ \ \ \ \ \ \    
+ 2 \left(\frac{\Delta v}{v_{\rm c,i}}\right) \frac{\sin{\theta} 
\left(\sin{\left(f_{\rm i}-\phi\right)} - e_{\rm i} \sin{\phi} \right)}
{\sqrt{1-e_{\rm i}^2}}\bigg]^{-1}.
\label{af}
\end{equation}

Equation (\ref{af}) reveals important dynamical traits. First, the expression is tractable enough so that one can deduce by inspection that the minimum value of $a_{\rm f}>0$ occurs when the small object is subject to a coplanar, anti-tangential kick $\left( \theta=\pi/2, \phi=-\pi/2 \right)$ while the body resides at its orbital pericentre $\left(f_{\rm i}=0\right)$. In this case, one obtains

\[
a_{\rm f}|_{\left(f_{\rm i}=0, \, \theta=\frac{\pi}{2}, \, \phi=-\frac{\pi}{2}\right)}
\]

\begin{equation}
= 
a_{\rm i} \left[  
1 - \left( \frac{\Delta v}{v_{\rm c,i}} \right)^2
+ 2 \left( \frac{\Delta v}{v_{\rm c,i}} \right) \sqrt{\frac{1+e_{\rm i}}{1-e_{\rm i}}}
\right]^{-1}.
\label{almmina}
\end{equation}

Now consider the behaviour of equation (\ref{almmina}) as a function of $\Delta v/v_{\rm c,i}$. For no kick at all, $a_{\rm f} = a_{\rm i}$. Then, as $\Delta v$ increases, $a_{\rm f}$ decreases, until reaching a local minimum at $\Delta v/v_{\rm c,i} = \sqrt{(1+e_{\rm i})/(1-e_{\rm i})}$. At this value,

\begin{equation}
{\rm min}\left(a_{\rm f} \right) = \frac{1}{2}q_{\rm i}.
\label{minminaf}
\end{equation}

\noindent{}Increasing $\Delta v/v_{\rm c,i}$ further increases $a_{\rm f}$. At the value of $\Delta v/v_{\rm c,i} = 2\sqrt{(1+e_{\rm i})/(1-e_{\rm i})}$, then we have come full circle, where again $a_{\rm f} = a_{\rm i}$. Increasing $\Delta v/v_{\rm c,i}$ even further then {\it increases} the small object's initial semi-major axis. Eventually, $\Delta v/v_{\rm c,i}$ becomes high enough to eject the small object, so that its semi-major axis becomes negative\footnote{A few examples of these critical values for escape for a coplanar, anti-tangential kick are, for $e_{\rm i} = 0$, $\Delta v/v_{\rm c,i} > 1 +\sqrt{2} \approx 2.4$; for $e_{\rm i} = 0.5$, $\Delta v/v_{\rm c,i} > 2 +\sqrt{3} \approx 3.7$; for $e_{\rm i} = 0.9$, $\Delta v/v_{\rm c,i} > 2\sqrt{5} +\sqrt{19} \approx 8.8$.}.

This behaviour is reflective of the more general case with arbitrary $f_{\rm i}$, $\theta$ and $\phi$: i.e. there is a finite range of kick values which actually decrease the small object's initial semi-major axis. Further, equation (\ref{minminaf}) illustrates that the initial semi-major axis can never be decreased by more than half of its original pericentre value. This decrease would not be sufficient to propel small objects with $q_{\rm i} > 200$~au into the perturbation zone without the small object's eccentricity also changing.

\section{How kicks stretch orbits}

Given the restriction on the reduction of $a_{\rm f}$ (equation \ref{minminaf}), I now consider how $q_{\rm i}$ can be lowered. Concurrently, I also consider how $Q_{\rm i}$ will change, given the escape constraint provided by $r_{\rm esc}$.

\subsection{Expressions for pericentre and apocentre}

To derive expressions for the post-blast pericentre distance $q_{\rm f}$ and apocentre distance $Q_{\rm f}$, I need the equation for $e_{\rm f}$ \citep{jacetal2014}. This relation may be expressed as

\begin{equation}
e_{\rm f} = 
\left[1 - \left(1-e_{\rm i}^2\right) K \frac{a_{\rm i}}{a_{\rm f}} \right]^{1/2} 
,
\label{ef}
\end{equation}

\noindent{}where

\[
K = \left[1 -  \left(\frac{\Delta v}{v_{\rm c,i}}\right) 
\frac{\sqrt{1-e_{\rm i}^2} \sin{\theta} \sin{\left(f_{\rm i}-\phi\right)} }{1+e_{\rm i}\cos{f_{\rm i}}}  \right]^2
\]

\begin{equation}
\ \ \ \ \ \ \ \
+ 
\left(\frac{\Delta v}{v_{\rm c,i}}\right)^2 
\frac{\left(1-e_{\rm i}^2 \right) \cos^2{\theta}}{\left(1+e_{\rm i}\cos{f_{\rm i}}\right)^2}
> 0.
\label{Kexp}
\end{equation}

\noindent{}Then,

\begin{equation}
q_{\rm f} = a_{\rm f} 
\left\lbrace 1 - \left[ 1 - \left(1-e_{\rm i}^2\right) K \frac{a_{\rm i}}{a_{\rm f}} \right]^{1/2} \right\rbrace,
\label{qf}
\end{equation}

\begin{equation}
Q_{\rm f} = a_{\rm f} 
\left\lbrace 1 + \left[ 1 - \left(1-e_{\rm i}^2\right) K \frac{a_{\rm i}}{a_{\rm f}} \right]^{1/2} \right\rbrace.
\label{bigQf}
\end{equation}

\subsection{Minimum value of pericentre}

A then relevant question is: does there exist a minimum value of $q_{\rm f}$, like for $a_{\rm f}$ (equation \ref{minminaf})? The answer is no. As proof, I show an example of a kick producing the extreme case of $q_{\rm f} = 0$, independent of $a_{\rm i}$. In this case, one solution of equation (\ref{qf}) is

\[
\left(\frac{\Delta v}{v_{\rm c,i}}\right)_{q_{\rm f} = 0} = \frac{1}{\sqrt{1-e_{\rm i}^2}}
\bigg[ \sin{\theta} \left( \sin{\left(f_{\rm i} - \phi \right)} - e_{\rm i} \sin{\phi}   \right)
\]

\begin{equation}
\ \ \ \ \ \ \ \ \ \ \ \ \ \ 
+\sqrt{1 - e_{\rm i}^2 + \sin^2{\theta} \left( \sin{\left(f_{\rm i} - \phi\right)} - 
e_{\rm i} \sin{\phi} \right)^2} \bigg].
\label{qf0}
\end{equation}

In equation (\ref{qf0}), an anti-radial kick at the apocentre ($f_{\rm i}=\pi$, $\phi=0$) satisfies $\Delta v/v_{\rm c,i} = 1$. Hence, indeed, even when a small body is at its apocentre, a supernova can alter its orbit so that $q_{\rm f}$ intersects the white dwarf.

Hence, because there is no minimum value of $q_{\rm f}$ but there is one for $a_{\rm f}$, I obtain a condition on $e_f$ when the supernova kick allows for $q_f < r_{\rm pert}$. The condition is, as long as $r_{\rm pert} < q_{\rm i}$,

\begin{equation}
e_{\rm f} > 1 - \frac{2 r_{\rm pert}}{a_{\rm i}\left(1 - e_{\rm i}\right)}.
\end{equation}

\subsection{Solution for kick}

Now I solve equations (\ref{qf}) and (\ref{bigQf}) for general $q_{\rm f}$ and $Q_{\rm f}$ in terms of $\Delta v/v_{\rm c,i}$. These are quartic equations in $\Delta v/v_{\rm c,i}$, which reflects the behaviour shown in Section 4 that there is a range of $\Delta v/v_{\rm c,i}$ values that allows for $q_{\rm f}<q_{\rm i}$. 

Two potentially real solutions for $\Delta v/v_{\rm c,i}$ for {\it both} of the equations are

\begin{equation}
\left(\frac{\Delta v}{v_{\rm c,i}}\right)_{1,2}
=
\frac
{-2A + 2A e_{\rm i}^2 + 2B W^2 \pm \sqrt{D}}
{2 \left[W^2 - C + e_{\rm i}^2C \right]}
\label{solvratio}
\end{equation}

\noindent{}where

\begin{equation}
A = \frac
{\sqrt{1-e_{\rm i}^2} \sin{\theta} \sin{\left(f_{\rm i} - \phi\right)}}
{1 + e_{\rm i} \cos{f_{\rm i}}},
\end{equation}

\begin{equation}
B = \frac
{\sin{\theta} \left( \sin{\left(f_{\rm i} - \phi\right)} - e_{\rm i} \sin{\phi}  \right)}
{\sqrt{1-e_{\rm i}^2} },
\end{equation}

\begin{equation}
C = \frac
{\left(1 - e_{\rm i}^2 \right) \left(\cos^2{\theta} + 
\sin^2{\theta}\sin^2{\left(f_{\rm i} - \phi\right)}\right)}
{\left(1 + e_{\rm i}\cos{f_{\rm i}} \right)^2 },
\end{equation}

\noindent{}and

\[
D = 4\left[A - A e_{\rm i}^2 - B W^2 \right]^2
\]

\begin{equation}
\ \ \ \ \ \
+4 
\left[W^2 - C + e_{\rm i}^2 C\right]
\left[1 - e_{\rm i}^2 - 2 W + W^2 \right],
\label{discgen}
\end{equation}

\noindent{}such that

\begin{equation}
W = \frac{q_{\rm f}}{a_{\rm i}} \ \ {\rm or} \ \ W = \frac{Q_{\rm f}}{a_{\rm i}}.
\end{equation}

%Further, equation (\ref{qf}) admits two more solutions:

%\begin{equation}
%\left(\frac{\Delta v}{v_{\rm c,i}}\right)_{3,4}
%=
%B \pm \sqrt{1 + B^2}.
%\end{equation}

In order to progress analytically, I note that for the purposes of this investigation, $q_{\rm f}/a_{\rm i}$ is often very small ($\ll 10^{-2}$) and always must be less than 2.3 per cent. This small $q_{\rm f}/a_{\rm i}$ regime constrains the relevant supernova geometry (Section 6). These constraints will then be applied to determine the allowable kick magnitudes (Section 7), when I will need to revisit the solutions of $\Delta v$ in terms of $Q_f$.

First though, in the limit of small $q_{\rm f}/a_{\rm i}$, the discriminant $D$ usefully reduces to

\[
D_{({\rm small} \ q_{\rm f}/a_{\rm i})} \approx \left[ \frac{2\left(1 - e_{\rm i}^2 \right)}{1 + e_{\rm i}\cos{f_{\rm i}}} \right]^2
\]

\begin{equation}
\ \ \ \,
\times 
\left[
\frac{2q_{\rm f}}{a_{\rm i}} \sin^2{\theta} \sin^2{\left\lbrace \left(f_{\rm i} - \phi\right) \right\rbrace}
-
\left( 
1 - e_{\rm i}^2 - \frac{2q_{\rm f}}{a_{\rm i}}
\right)
\cos^2{\theta}
\right].
\label{discred}
\end{equation}

\section{Constraining the supernova geometry}

\subsection{Offset from coplanarity}

Now I use equation (\ref{discred}) to determine the probability that the geometry of a randomly oriented supernova will thrust a small object's pericentre inwards to within $r_{\rm pert}$. The equation provides a restriction on the range of possible values of $\theta$:

\[
-\left[1 + \frac
{2\frac{q_{\rm f}}{a_{\rm i}} \sin^2{\left(f_{\rm i} - \phi\right)}}
{1 - e_{\rm i}^2 - 2\frac{q_{\rm f}}{a_{\rm i}}}   
\right]^{-\frac{1}{2}}
\]

\begin{equation}
\le
\sin{\theta}
\le
\left[1 + \frac
{2\frac{q_{\rm f}}{a_{\rm i}} \sin^2{\left(f_{\rm i} - \phi\right)}}
{1 - e_{\rm i}^2 - 2\frac{q_{\rm f}}{a_{\rm i}}}   
\right]^{-\frac{1}{2}}.
\label{condtheta}
\end{equation}

Equation (\ref{condtheta}) illustrates that a radial or anti-radial kick ($\phi = f_{\rm i}$ or $\phi = f_{\rm i} + \pi$) allows for small $q_{\rm f}/a_{\rm i}$ only when the kick is exactly co-planar ($\theta = \pi/2$). Alternatively, the allowable range of $\theta$ is maximised when the kick is anti-tangential ($\phi = f_{\rm i} - \pi/2$), and is bounded by

\begin{equation}
{\rm min}\left(\theta\right) 
= {\rm ArcSin}
\left\lbrace T \right\rbrace
,
\label{mintheta}
\end{equation}

\begin{equation}
{\rm max}\left(\theta\right) 
=\pi + {\rm ArcSin}
\left\lbrace -T \right\rbrace
,
\label{maxtheta}
\end{equation}

\noindent{}where

\[
T = \sqrt{
\frac
{\left(1 - e_{\rm i}^2\right)  \left[\left(1 - \frac{q_{\rm f}}{a_{\rm i}}\right)^2 - e_{\rm i}^2 \right]}
{\left(1 - e_{\rm i}^2\right)^2 - \left(\frac{q_{\rm f}}{a_{\rm i}}\right)^2 
\left(1 + e_{\rm i} \cos{f_{\rm i}}  \right)^2
}
}
\approx \sqrt{1 - \frac
{2\frac{q_{\rm f}}{a_{\rm i}}}
{1 - e_{\rm i}^2}
}
.
\]

\begin{equation}
\label{eqforT}
\end{equation}

The first expression in equation (\ref{eqforT}) is the result of applying $\phi = f_{\rm i} - \pi/2$ to equation (\ref{discgen}), and shows that $T$ is generally a function of $f_{\rm i}$, but that this dependence vanishes in the small $q_{\rm f}/a_{\rm i}$ approximation.

\subsection{Probability of achieving required geometry}

I now assume that the supernova blast wave is isotropically distributed on the sky such that $\theta = \cos^{-1}\left(2w-1\right)$, where $w$ is uniformly sampled from 0 to 1, and $\phi$ is uniformly sampled from 0 to $2\pi$. Then, I obtain an expression for $\phi(\theta)$ from equation (\ref{condtheta}) to determine the probability of a blast propelling a small body pericentre inwards as

\[
\mathcal{P}(\Delta\theta,\Delta\phi) = 
\]

\[
\int_{{\rm min}\left(\theta\right) }^{\frac{\pi}{2}}
\left\lbrace
\frac{1}{2}
-
\frac{1}{\pi}
\left[
{\rm ArcSin}{
\left(
\cot{\theta}
\sqrt{\frac{1 - e_{\rm i}^2}{2 \frac{q_{\rm f}}{a_{\rm i}}} - 1}
\right)
}
\right]
\right\rbrace
\sin{\theta}
d\theta
.
\]

\begin{equation}
\label{exactP}
\end{equation}

Equation (\ref{exactP}) is exact subject to the small $q_{\rm f}/a_{\rm i}$ approximation, and can be readily computed numerically. However, obtaining an analytical solution to the integral is difficult, as is representing the integrand by a manageable series expansion.

Hence, in order to obtain a more convenient numerical estimate, I assume $\mathcal{P}(\Delta\theta,\Delta\phi) \approx \mathcal{P}(\Delta\theta)\mathcal{P}(\Delta\phi)$. Then, $\mathcal{P}(\Delta\theta) \propto \sin{\theta}d\theta$, and

\begin{equation}
\mathcal{P}(\Delta\theta) = \cos{\left\lbrace{\rm ArcSin}\left\lbrace T \right\rbrace\right\rbrace}
=\sqrt{1 - T^2}
\approx 
\sqrt{\frac
{2\frac{q_{\rm f}}{a_{\rm i}}}
{1 - e_{\rm i}^2}
}
.
\label{Ptheta}
\end{equation}

%%%%%%%%%%%%%%%% Figure
\begin{figure*}
\includegraphics[width=17.0cm]{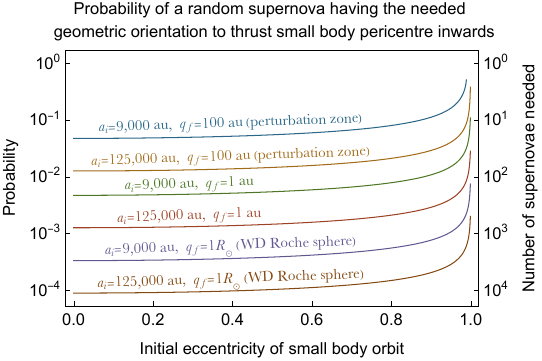}
\caption{
Probabilities of supernova blast directions which propel orbital pericentres of small bodies inwards, from equation (\ref{exactP}) or (\ref{Ptotal}); both equations produce results that are visually indistinguishable on the plot. The $x$-axis is represented by the parameter $e_{\rm i}$, and the small body's initial semi-major axis and final pericentre are given by the different curves. The values given on the $y$-axes assume that the supernova explosions are randomly oriented on the sky. The two values of $a_{\rm i}$ sampled for the curves correspond to the escape boundary of the system, $r_{\rm esc} =$ 125,000~au, and the region within which the impulse approximation begins to break down, $r_{\rm imp} =$ 9,000~au. This figure demonstrates that less than 100 supernova would be needed before at least one generates a blast which can geometrically propel exo-Oort cloud-like planetary debris into the perturbation zone. For highly eccentric orbits near $r_{\rm imp}$, only a few such supernovae would be needed.
}
\label{ProbFig}
\end{figure*}
%%%%%%%%%%%%%%%% Figure

Equation (\ref{Ptheta}) illustrates explicitly that the allowable range of $\theta$ decreases as the small object's orbit becomes more circular. In contrast, for highly eccentric initial orbits, a substantially greater array of values are $\theta$ allowed. 

In order to estimate $\mathcal{P}(\Delta\phi)$, I compare areas on a plot of $\theta$ vs $(f_{\rm i} - \phi)$ by computing the series-expanded integral of the curve given by (from equation \ref{condtheta})

\[
\theta \approx {\rm ArcSin}\left\lbrace  
\left[1 + \frac
{2\frac{q_{\rm f}}{a_{\rm i}} \sin^2{\left(f_{\rm i} - \phi\right)}}
{1 - e_{\rm i}^2 - 2\frac{q_{\rm f}}{a_{\rm i}}}   
\right]^{-\frac{1}{2}} \right\rbrace
\]

\begin{equation}
\ \ 
\approx \frac{\pi}{2} - \sin{\left(f_{\rm i} - \phi\right)} 
        \sqrt{\frac{2 \frac{q_{\rm f}}{a_{\rm i}}}{1-e_{\rm i}^2}},
\end{equation}

\noindent{}yielding the final result:

\[
\mathcal{P}(\Delta\theta,\Delta\phi)
\sim k
\mathcal{P}(\Delta\theta)
\left[
\frac
{\frac{\pi^2}{4} - \sqrt{\frac{2 \frac{q_{\rm f}}{a_{\rm i}} }{1-e_{\rm i}^2}}  - \frac{\pi}{2}{\rm min}\left(\theta\right)}
{\frac{\pi}{2} \left(\frac{\pi}{2} - {\rm min}\left(\theta\right) \right)}
\right]
\]

\[
= k
\sqrt{\frac{2 \frac{q_{\rm f}}{a_{\rm i}} }{1-e_{\rm i}^2}}
\left[
1
-
\frac
{\sqrt{\frac{2 \frac{q_{\rm f}}{a_{\rm i}} }{1-e_{\rm i}^2}}}
{\frac{\pi}{2} \left(\frac{\pi}{2} - {\rm ArcSin}\left\lbrace \sqrt{1 - \frac
{2\frac{q_{\rm f}}{a_{\rm i}}}
{1 - e_{\rm i}^2}
}\right\rbrace \right)}
\right]
\label{Ptotal0b}
\]

\begin{equation}
\approx 
k
\left(1 - \frac{2}{\pi} \right)
\sqrt{\frac{2 \frac{q_{\rm f}}{a_{\rm i}} }{1-e_{\rm i}^2}},
\label{Ptotal}
\end{equation}

\noindent{}where $k$ is a numerical factor that corrects for the approximation $\mathcal{P}(\Delta\theta,\Delta\phi) \approx \mathcal{P}(\Delta\theta)\mathcal{P}(\Delta\phi)$. I find through numerical tests (see Section 8) that $k\approx 0.87$ is largely independent of both $e_{\rm i}$ and $q_{\rm f}/a_{\rm i}$, and yields correct answers to within a few per cent. Given that the probabilities $\mathcal{P}(\Delta\theta,\Delta\phi)$ span many orders of magnitude depending on the values of $q_{\rm f}/a_{\rm i}$, this level of error is effectively negligible.

I plot equation (\ref{Ptotal}) in Fig. \ref{ProbFig} for a variety of configurations of $a_{\rm i}$ and $q_{\rm f}$ as a function of $e_{\rm i}$. These configuration choices bound extreme scenarios, and demonstrate that only a few randomly oriented supernovae would be needed to generate inwards pericentre shifts into the perturbation zone for high $e_{\rm i}$ and $a_{\rm i}\sim 10^4$~au, whereas over $10^3$ supernovae would be needed to alter a small body orbit so that its pericentre intersects the white dwarf's Roche sphere.

\section{Constraining the kick magnitude}

Having computed the probability of a supernova blast to reside in a geometric configuration that allows for an inward pericentre push, I now seek to analytically bound the possible values of $\Delta v$, a quantity which is always assumed to be positive.

\subsection{Kick needed to achieve inwards pericentre thrust}

First, consider the minimum possible kick that will change the pre-blast orbit such that the post-blast orbit satisfies $q_{\rm f} \le r_{\rm pert}$. At either $\theta = {\rm min}(\theta)$ or $\theta = {\rm max}(\theta)$, the kick would need to be anti-tangential ($\phi = f_{\rm i} - \pi/2$). As a result, from equations (\ref{solvratio}) and (\ref{discgen}), $D=0$, and 

\[
{\rm min}\left(\frac{\Delta v}{v_{\rm c,i}}\right)_{\rm pert} = 
\]

\begin{equation}
\ \ \ 
\left(1 + e_{\rm i} \cos{f_{\rm i}} \right)
\sqrt{ 
\frac
{\left(1 - \frac{r_{\rm pert}}{a_{\rm i}} \right)^2 - e_{\rm i}^2}
{\left(1 - e_{\rm i}^2 \right)^2 - \left( \frac{r_{\rm pert}}{a_{\rm i}} \right)^2 
\left(1 + e_{\rm i} \cos{f_{\rm i}} \right)^2}
}.
\label{minvlimitexact}
\end{equation}

\noindent{}In the limit of small $r_{\rm pert}/a_{\rm i}$, this condition reduces to:

\[
{\rm min}\left(\frac{\Delta v}{v_{\rm c,i}}\right)_{\rm pert} \approx 
\frac{1 + e_{\rm i}\cos{f_{\rm i}}}{\sqrt{1-e_{\rm i}^2}}
-
\frac{1 + e_{\rm i}\cos{f_{\rm i}}}{1-e_{\rm i}^2}
\sqrt{\frac{2r_{\rm pert}}{a_{\rm i}}}
\]

\begin{equation}
\ \ \ \ \ \ \ \ \ \ \ \ \ \ \ \ \ \ \ \ \,
\approx 
\frac{1 + e_{\rm i}\cos{f_{\rm i}}}{\sqrt{1-e_{\rm i}^2}}.
\label{minvlimit}
\end{equation}

%%%%%%%%%%%%%%%% Figure 
\begin{figure*}
%\centerline{\Huge \underline{Planet}}
\centerline{}
\centerline{}
\centerline{
\includegraphics[width=9.0cm]{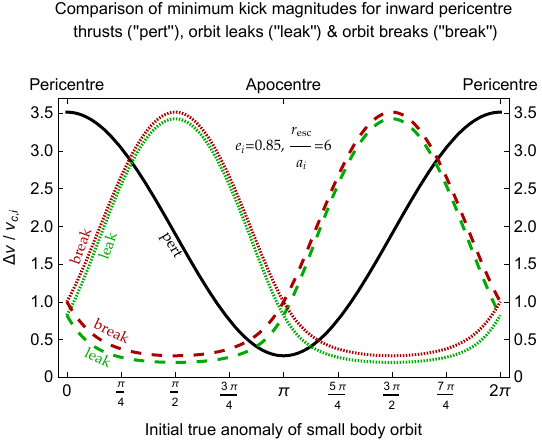}
\includegraphics[width=9.0cm]{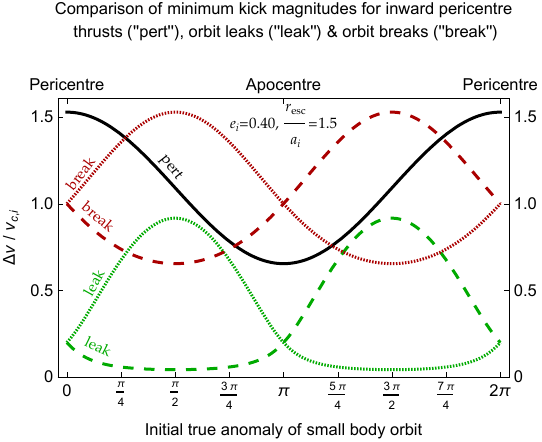}
}
\caption{
Comparison of minimum kick formulations. On both plots, the solid black curve (``pert"; equation \ref{minvlimit}) is the minimum kick required to change the orbit of a small body so that its post-blast pericentre is within the perturbation zone ($q_{\rm f}\le100$~au). The green curves labelled ``leak" indicate all branches of equation (\ref{minleak}) which yield positive kick values; the smaller of these indicates the minimum kick needed to generate $Q_{\rm f} > r_{\rm esc}$ with a post-blast osculating elliptical orbit. The red curves labelled ``break" indicate all branches of equation (\ref{minbreak}) which yield positive kick values; the smaller of these indicates the minimum kick needed to convert the pre-blast elliptical orbit into a post-blast hyperbolic one. In the left panel, the ``pert" curve is below all others in a narrow region around $f_{\rm i} = \pi$, whereas in the right panel, no such region exists.
}
\label{KicksFig}
\end{figure*}
%%%%%%%%%%%%%%%% Figure 

%%%%%%%%%%%%%%%% Figure
\begin{figure}
\includegraphics[width=8.5cm]{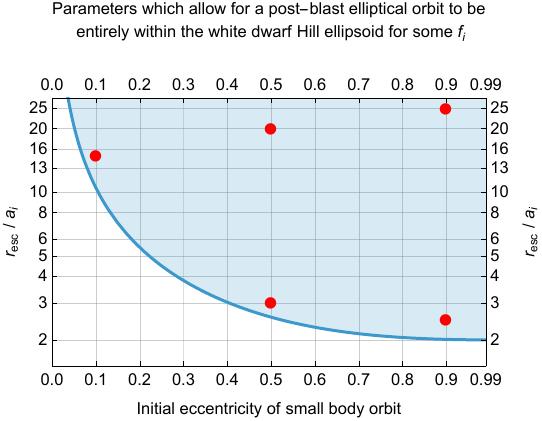}
\caption{
Pre-blast orbital properties ($a_{\rm i}$, $e_{\rm i}$) that can allow for repeated post-blast pericentre passages of the small body in the perturbation zone, for some range (not specified here) of $f_{\rm i}$. The shaded region, above the curve (equation \ref{pertcond}), indicates the combinations of initial eccentricities and semi-major axes which {\it may} allow for these repeated pericentre passages to occur. The region below the curve indicates where the post-blast orbit will always leak or break, limiting the number of passages of the small body in the perturbation zone to at most one. The red dots have been chosen to generate approximate but specific limiting values of the maximum affected size of minor bodies in Table \ref{TableNvalFig}.
}
\label{KeepBoundFig}
\end{figure}
%%%%%%%%%%%%%%%% Figure

\subsection{Orbit contortion possibilities}

In order to place the magnitude of this kick into the context of the effect on the orbit, I outline three possible scenarios:

\begin{enumerate}

\item The kick magnitude bounded by equation (\ref{minvlimit}) stretches the orbit in such a way that both $q_{\rm f} \le r_{\rm pert}$ and $Q_{\rm f} < r_{\rm esc}$. As a result, the entire orbit remains within the Hill ellipsoid of the white dwarf, and the small body will repeatedly enter the perturbation zone (subject to inter-blast exo-Oort cloud dynamics on secular timescales).

\,

\item The kick magnitude given by equation (\ref{minvlimit}) is already high enough to stretch the orbit so that both $q_{\rm f} \le r_{\rm pert}$ and $Q_{\rm f} \ge r_{\rm esc}$. In this case, although the osculating ellipse of the orbit remains well-defined, the small body will leave the system on its approach to apocentre. Therefore, the body may or may not enter the perturbation zone just once, depending on the value of $f_{\rm f}$.

\,

\item The kick magnitude given by equation (\ref{minvlimit}) is high enough to break the elliptical orbit and convert it into a hyperbolic one (where $a_{\rm f} < 0$; see equation \ref{af}). The small body, despite now being on a hyperbolic orbit, may still pass through the perturbation zone, but only once.

\end{enumerate}

Which of the three scenarios above occurs depends on the values of $\theta$, $\phi$, $a_{\rm i}$, $e_{\rm i}$ and $f_{\rm i}$, subject to the restrictions established on $\theta$ and $\phi$ in the last section.

\subsection{Kick needed to generate orbital leaking}

In order to constrain when scenarios (i) or (ii) occur, I set $W = Q_{\rm f}/a_{\rm i}$ and $Q_{\rm f} = r_{\rm esc}$ in equation (\ref{solvratio}), and then use the geometrical constraints from the last section on the values of $\theta$ and $\phi$. In particular, the minimum value of the kick magnitude that will cause the small body to ``leak" out of the system occurs when no component of the kick is out of the plane of the orbit ($\theta = \pi/2$). In this case, the kick needs to be radial or anti-radial ($\phi=f_{\rm i}, \phi=f_{\rm i}+\pi$). Inserting these values into equations (\ref{solvratio}) and (\ref{discgen}) gives

\[
{\rm min}\left(\frac{\Delta v}{v_{\rm c,i}}\right)_{\rm leak} = 
\]

\[
\frac
{\pm e_{\rm i}\sin{f_{\rm i}} \pm
\sqrt{e_{\rm i}^2 \sin^2{f_{\rm i}} + \left(\frac{a_{\rm i}}{r_{\rm esc}} \right)^2
\left(1 - e_{\rm i}^2 \right)
\left[\left(1 - \frac{r_{\rm esc}}{a_{\rm i}} \right)^2 - e_{\rm i}^2 \right]  }
}
{\sqrt{1-e_{\rm i}^2}}
.
\]

\begin{equation}
\label{minleak}
\end{equation}

\noindent{}In the first sign choice, the upper sign corresponds to anti-radial kicks and the lower sign corresponds to radial kicks. For anti-radial kicks, equation (\ref{minleak}) is guaranteed to generate positive kick values if the initial orbit is within the Hill ellipsoid. Further, in this case, the upper sign of the second sign choice can be assumed.

\subsection{Kick needed to generate orbital breaking}

In order to determine when scenario (iii) is applicable, I obtain a minimum kick velocity that ``breaks" the orbit by equating the term in brackets in equation (\ref{af}) to 0, again for coplanar ($\theta = \pi/2$) and radial or anti-radial ($\phi=f_{\rm i}, \phi=f_{\rm i}+\pi$) kicks. The result is:

\begin{equation}
{\rm min}\left(\frac{\Delta v}{v_{\rm c,i}}\right)_{\rm break} = 
\frac
{\pm e_{\rm i}\sin{f_{\rm i}} \pm \sqrt{1 - e_{\rm i}^2\left(1 - \sin^2{f_{\rm i}}\right)}}
{\sqrt{1-e_{\rm i}^2}}
.
\label{minbreak}
\end{equation}

\noindent{}Again, in the first sign choice, the upper sign corresponds to anti-radial kicks and the lower sign corresponds to radial kicks. 

\subsection{Kick comparisons}

It must be the case that for a given geometrical blast configuration,  

\begin{equation}
{\rm min} \left(\frac{\Delta v}{v_{\rm c,i}}\right)_{\rm leak} <
{\rm min} \left(\frac{\Delta v}{v_{\rm c,i}}\right)_{\rm break}.
\end{equation}

\noindent{}Indeed, this relation can be confirmed through a comparison of equations (\ref{minleak}) and (\ref{minbreak}).

As a specific example, for initially circular orbits,

\[
{\rm min}\left(\frac{\Delta v}{v_{\rm c,i}}\right)_{{\rm leak},e_{\rm i}=0} = 
\left|\frac{a_{\rm i}}{r_{\rm esc}} - 1 \right|
\]

\begin{equation}
<
{\rm min}\left(\frac{\Delta v}{v_{\rm c,i}}\right)_{{\rm break},e_{\rm i}=0} = 1,
\end{equation}

\noindent{}and min$(\Delta v/v_{\rm c,i})_{{\rm pert},e_{\rm i}=0} = 1$. Hence, in order for a small object on an initially circular orbit to reach the perturbation zone, the supernova blast must break the ellipticity of the orbit and place the small object on a hyperbolic orbit.

In order to illustrate the interplay amongst these kick limits, I plot two specific cases in Fig. \ref{KicksFig} for all solution branches where $\Delta v > 0$. The left panel demonstrates an example where min$(\Delta v/v_{\rm c,i})_{\rm pert} <$ min$(\Delta v/v_{\rm c,i})_{\rm leak}$ for some $f_{\rm i}$ range, whereas in the right panel this relation is not satisfied for any $f_{\rm i}$.

%%%%%%%%%%%%%%%% Figure
\begin{figure*}
\includegraphics[width=17.0cm]{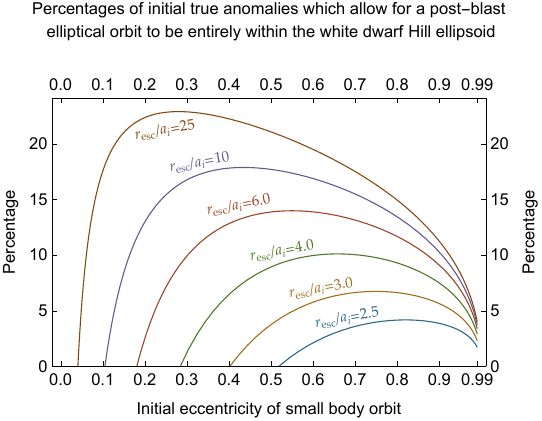}
\caption{
The effective size of the region around the initial apocentre where the ``pert" curves are lower than all others (e.g. see left panel of Fig. \ref{KicksFig}), enabling for a bounded post-blast orbit where repeated pericentre passages occur. Plotted here is equation (\ref{frangefrac}). This figure illustrates that in all cases, the allowable $f_{\rm i}$ interval is under 23.3 per cent of the variable's entire range (see also equation \ref{maxFeq}).
}
\label{frangeFig}
\end{figure*}
%%%%%%%%%%%%%%%% Figure

\subsection{Conditions for repeated pericentre passages}

The applicable range of $f_{\rm i}$ which can allow for post-blast orbits to remain elliptic and entirely within the Hill ellipsoid is centred around $\pi$. This property is generally true: the small body must be near the apocentre pre-blast for there to be a chance of the object later experiencing repeated pericentre passages into the perturbation zone. 

In order to explore the conditions which allow for this property to hold, I compare equations (\ref{minvlimit}) and (\ref{minleak}) at $f_{\rm i} = \pi$, yielding the following condition

\[
\left(\frac{r_{\rm esc}}{a_{\rm i}}\right)_{{\rm ``pert"<``leak"}}
\gtrsim
\left(
\frac{1+e_{\rm i}}{2e_{\rm i}}
\right)
\left[
1 + \sqrt{1 - 2e_{\rm i} \left(1 - e_{\rm i} \right)}
\right]
,
\]

\begin{equation}
\label{pertcond}
\end{equation}

I plot equation (\ref{pertcond}) in Fig. \ref{KeepBoundFig}. The figure illustrates that if the small body's initial semi-major axis is greater than half of $r_{\rm esc}$, then the post-blast orbit will always leak or break. Further, initially eccentric orbits afford a much greater scope for retaining a confined, elliptic post-blast orbit than initially near-circular ones.

This retention is achieved for a range of $f_{\rm i}$ which may be quantified. This region can be approximated by expanding $f_{\rm i}$ about $\pi$ in the comparison of equations (\ref{minvlimit}) and (\ref{minleak}), giving

\begin{equation}
f_{\rm min} < f_{\rm i} < f_{\rm max},
\end{equation}

\noindent{}where

\begin{equation}
f_{\rm min} = 2\pi - f_{\rm crit},
\end{equation}

\begin{equation}
f_{\rm max} = f_{\rm crit},
\end{equation}

\noindent{}and

\[
f_{\rm crit} \approx \frac{e_{\rm i}\left(\pi + 1\right) - 1}{e_{\rm i}}
+
\frac{a_{\rm i}}{e_{\rm i} r_{\rm esc}}
\sqrt{\left(1 - e_{\rm i}^2 \right) 
\left[  
\left(1 - \frac{r_{\rm esc}}{a_{\rm i}} \right)^2 - e_{\rm i}^2
\right]}.
\]

\begin{equation}
\label{fcrit}
\end{equation}

Equation (\ref{fcrit}) is independent of $f_{\rm i}$, but only to first order in the expansion around the apocentre. The next term in the expansion is dependent on $f_{\rm i}$.

The fraction, $\mathcal{F}$, of true anomalies for which ${\rm min}(\Delta v)_{\rm pert} < {\rm min}(\Delta v)_{\rm leak}$ is then

\begin{equation}
\mathcal{F} = \frac{f_{\rm crit}}{\pi} - 1.
\label{frangefrac}
\end{equation}

I plot $\mathcal{F}$ in Fig. \ref{frangeFig} for six allowable $r_{\rm esc}/a_{\rm i}$ values (from Fig. \ref{KeepBoundFig}). By computing $\partial\mathcal{F}/\partial e_{\rm i}$, I can approximate the maximum value of $\mathcal{F}$, $\mathcal{F}_{\rm max}$, as a function of $r_{\rm esc}/a_{\rm i}$ by assuming both $e_{\rm i}^2 \ll 1$ and $r_{\rm esc}/a_{\rm i} \gg 1$. The result is 

\[
\mathcal{F}_{\rm max}
\approx
\frac{1}{\pi}
\left[
1+\frac{\sqrt{2}}{2}
\left\lbrace
\sqrt{\frac{r_{\rm esc}}{a_{\rm i}} - 4}
-
\left(1 - \frac{a_{\rm i}}{r_{\rm esc}} \right)^{-1}
\sqrt{\frac{r_{\rm esc}}{a_{\rm i}} - 2}
\right\rbrace
\right].
\]

\begin{equation}
\label{maxFeq}
\end{equation}

Overall, the maximum possible value of $\mathcal{F}_{\rm max}$, which occurs for the extreme value of the ratio $r_{\rm esc}/a_{\rm i} = 125$,000/4,500 = 27.8, is 23.3 per cent numerically and 23.0 per cent from equation (\ref{maxFeq}). This maximum value applies to all typical white dwarfs located at approximately the Solar neighbourhood distance from the Galactic centre.

\subsection{Post-blast true anomalies}

For bounded post-blast orbits, repeated pericentre passages into the perturbation zone will occur regardless of the value of $f_{\rm f}$. However, for leaky post-blast orbits (where $Q_{\rm f} > r_{\rm esc}$) the value of $f_{\rm f}$ will determine if the small object will enter the perturbation zone once, or not. Expressions for $f_{\rm f}$ from \cite{jacetal2014} can be written as

\begin{equation}
\cos{f_{\rm f}} = \frac{1}{e_{\rm f}} \left[K \left(1 + e_{\rm i} \cos{f_{\rm i}}\right) - 1\right],
\label{cosf}
\end{equation}

\[
\sin{f_{\rm f}} = \frac{\sqrt{K}}{e_{\rm f}} 
                  \left[e_{\rm i}\sin{f_{\rm i}} - \sqrt{1 - e_{\rm i}^2} 
                  \left( \frac{\Delta v}{v_{\rm c,i}} \right) \sin{\theta} \cos{\left(\phi - f_{\rm i}\right)}
                   \right].
\]

\begin{equation}
\label{sinf}
\end{equation}

If the supernova blast wave is treated as instantaneous, then at the moment of interaction, the small body's position will not change even though its osculating orbit will. This property can be verified by combining equations (\ref{dist}), (\ref{ef}), and (\ref{cosf}). Therefore, as long as the post-blast orbit is an ellipse, regardless of its osculating apocentre value, the condition that $\pi < f_{\rm f} < 2\pi$ will ensure that the small body will reach the perturbation zone. That condition in turn requires that $\sin{f_{\rm f}}<0$.

I argue that this condition would be satisfied for about half of an initial uniform ring of debris that shares the same orbit. Consider, for example, the value of $\sin{f_{\rm f}}$ subject to an anti-tangential kick at the geometric bounds of $\theta$:

\[
\sin{f_{\rm f}}|_{\left( \theta = {\rm min}\theta \ {\rm or } \ \theta = {\rm max}\theta, \ \phi = f_{\rm i} - \pi/2 \right)}
=
\left(\frac{e_{\rm i}}{e_{\rm f}} \sin{f_{\rm i}}  \right)
\]

\[
\times
\frac
{\sqrt{2 \frac{r_{\rm pert}}{a_{\rm i}} \left(\frac{\Delta v}{v_{\rm c,i}}\right)^2 
+
\left(
1
-
\left(\frac{\Delta v}{v_{\rm c,i}}\right)
\sqrt{1 - e_{\rm i}^2 - \frac{r_{\rm pert}}{a_{\rm i}}}
+ e_{\rm i}\cos{f_{\rm i}}
\right)^2
}}
{1+e_{\rm i} \cos{f_{\rm i}}}
\]

\begin{equation}
\approx \frac{e_{\rm i}}{e_{\rm f}} \sin{f_{\rm i}}
\left[
1 - \left(\frac{\Delta v}{v_{\rm c,i}}\right)
\frac{\sqrt{1-e_{\rm i}^2}}{1+e_{\rm i}\cos{f_{\rm i}}}
\right]
.
\end{equation}

\noindent{}Hence, given the minimum value of the kick from equation (\ref{minvlimit}), here $\sin{f_{\rm f}}$ would just take on the opposite sign of $\sin{f_{\rm i}}$.

\section{Numerical tests}

In order to test the accuracy of the analytics, I now perform a numerical exercise. I sample the entire range of potential supernova blast geometries ($\theta = \left[0,\pi\right], \phi = \left[0,2\pi\right]$) for different small body values of $a_{\rm i}$, $e_{\rm i}$ and $f_{\rm i}$, and different values of $r_{\rm pert}$. The blast geometries are isotropically distributed on the sky, with an equivalent number of values of $\theta$ and $\phi$ sampled for each test case; in total, I sample $1.0 \times 10^6$, $2.5 \times 10^5$ and $4.0 \times 10^6$ blast geometries, respectively.

\begin{itemize}

\item {\bf Test \#1}: I assume that the entire pre-blast small body orbit is within $r_{\rm imp}$ and $r_{\rm esc}$ such that $a_{\rm i} = $ 50,000~{\rm au}, $e_{\rm i} =$ 0.2, $q_{\rm i} = $ 40,000~{\rm au} and $Q_{\rm i} = $ 60,000~{\rm au}. The small body's initial position is given by an eccentric anomaly of $3\pi/2$, which corresponds to $f_{\rm i} = 4.51$~{\rm rad}, and the goal is to alter the object's pericentre to $q_{\rm f} \le 100$~{\rm au}. The numerical results reveal that the total percentage of blast geometries which allow for an inward pericentre thrust into the perturbation zone from the numerical simulation is 2.060 per cent; from equation (\ref{exactP}), this value is $\mathcal{P}(\Delta\theta,\Delta\phi) = 2.056$ per cent, and from equation (\ref{Ptotal}), this value is $\mathcal{P}(\Delta\theta,\Delta\phi) = 2.041$ per cent.

\,

\item {\bf Test \#2}: Here I assume that the initial pericentre of the small body is well within $r_{\rm imp}$, but the body is initially at its apocentre, such that $f_{\rm i} = \pi$, $a_{\rm i} = $ 9,000~{\rm au}, $e_{\rm i} =$ 0.90, $q_{\rm i} = $ 900~{\rm au}, and $Q_{\rm i} = $ 17,100~{\rm au}. I compute solutions for $q_{\rm f} \le 100$~{\rm au}. I find that the total percentage of blast geometries which allow for an inward pericentre thrust into the perturbation zone from the numerical simulation is 11.09 per cent; from equation (\ref{exactP}), this value is $\mathcal{P}(\Delta\theta,\Delta\phi) = 11.11$ per cent, and from equation (\ref{Ptotal}), this value is $\mathcal{P}(\Delta\theta,\Delta\phi) = 10.81$ per cent.

\,

\item {\bf Test \#3}:  Alternatively, here I assume that the initial osculating apocentre of the small body is well outside of $r_{\rm esc}$ but the body is initially at its pericentre, such that $f_{\rm i} = 0$, $a_{\rm i} = $ 125,000~{\rm au}, $e_{\rm i} =$ 0.90, $q_{\rm i} = $ 12,500~{\rm au}, and $Q_{\rm i} = $ 237,500~{\rm au}. Further, in this extreme test, I model a deep penetration into the perturbation zone so that $q_{\rm f} \le 1$~{\rm au}.  I find that the total percentage of blast geometries which allow for an inward pericentre thrust into this 1~au region from the numerical simulation is 0.2935 per cent; from equation (\ref{exactP}), this value is $\mathcal{P}(\Delta\theta,\Delta\phi) = 0.2921$ per cent, and from equation (\ref{Ptotal}), this value is $\mathcal{P}(\Delta\theta,\Delta\phi) = 0.2901$ per cent.

\end{itemize}

\begin{figure*}
\centerline{\Large \underline{Numerical tests (Test \#1: top row, Test \#2: middle row, Test \#3: bottom row)}}
\centerline{}
\centerline{
\includegraphics[width=9.0cm]{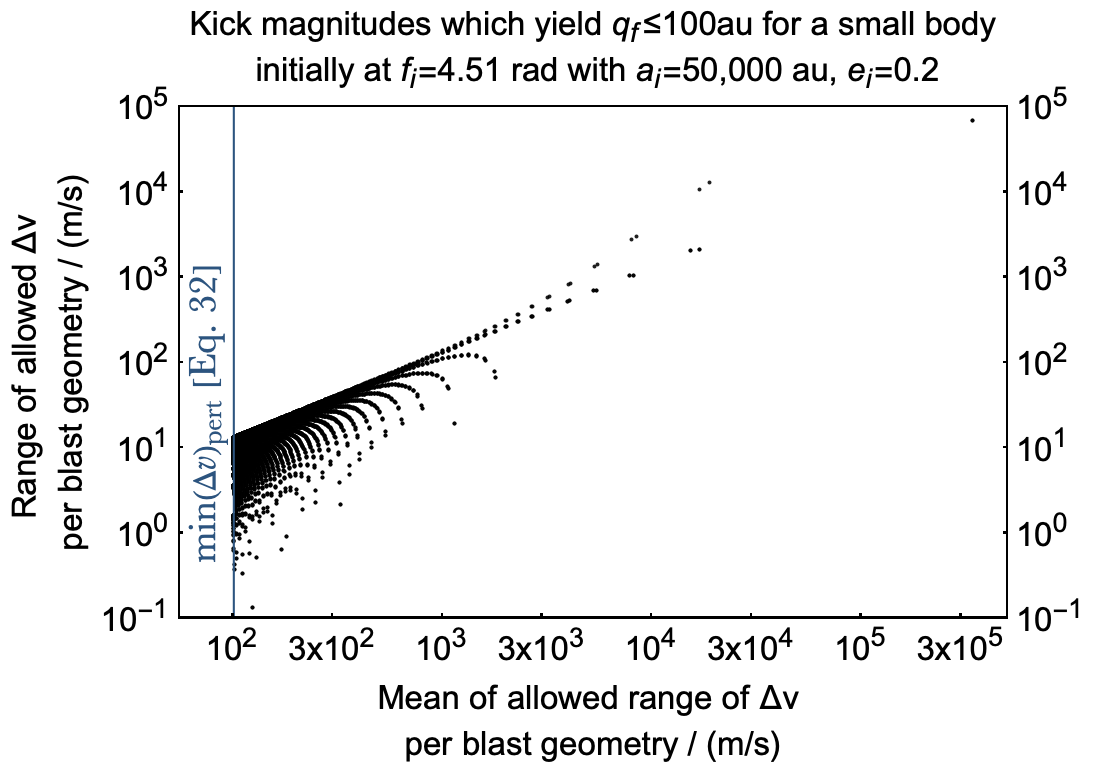}
\includegraphics[width=9.0cm]{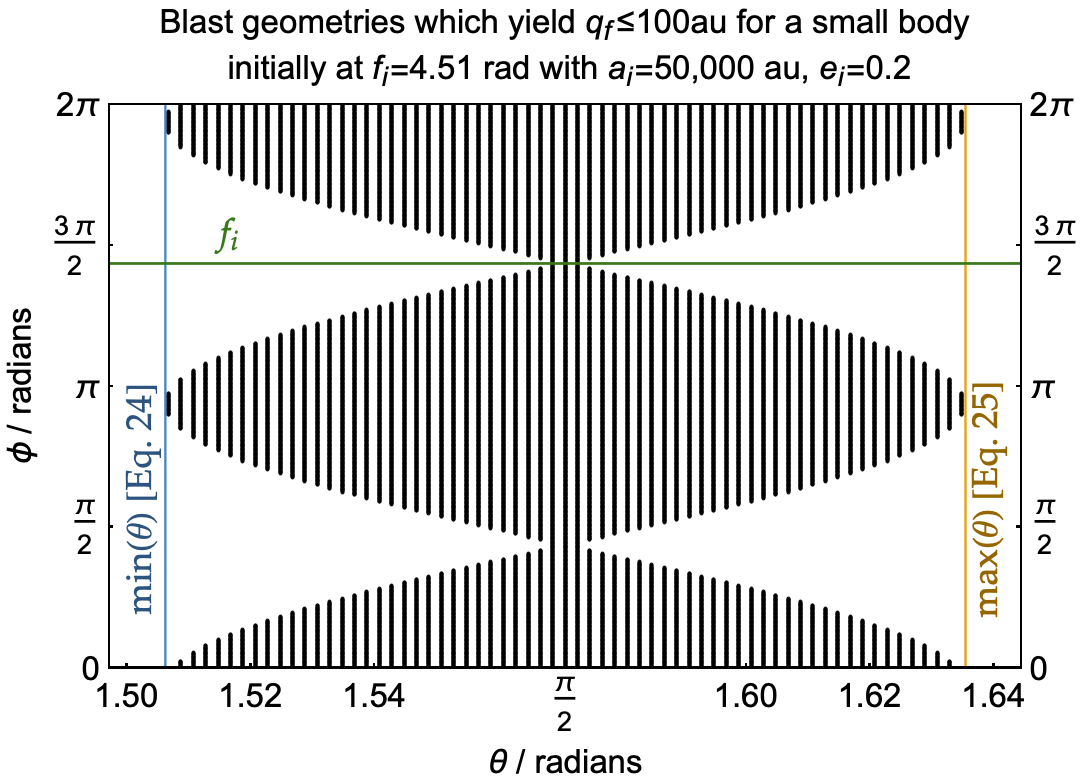}
}
\centerline{}
\centerline{
\includegraphics[width=9.0cm]{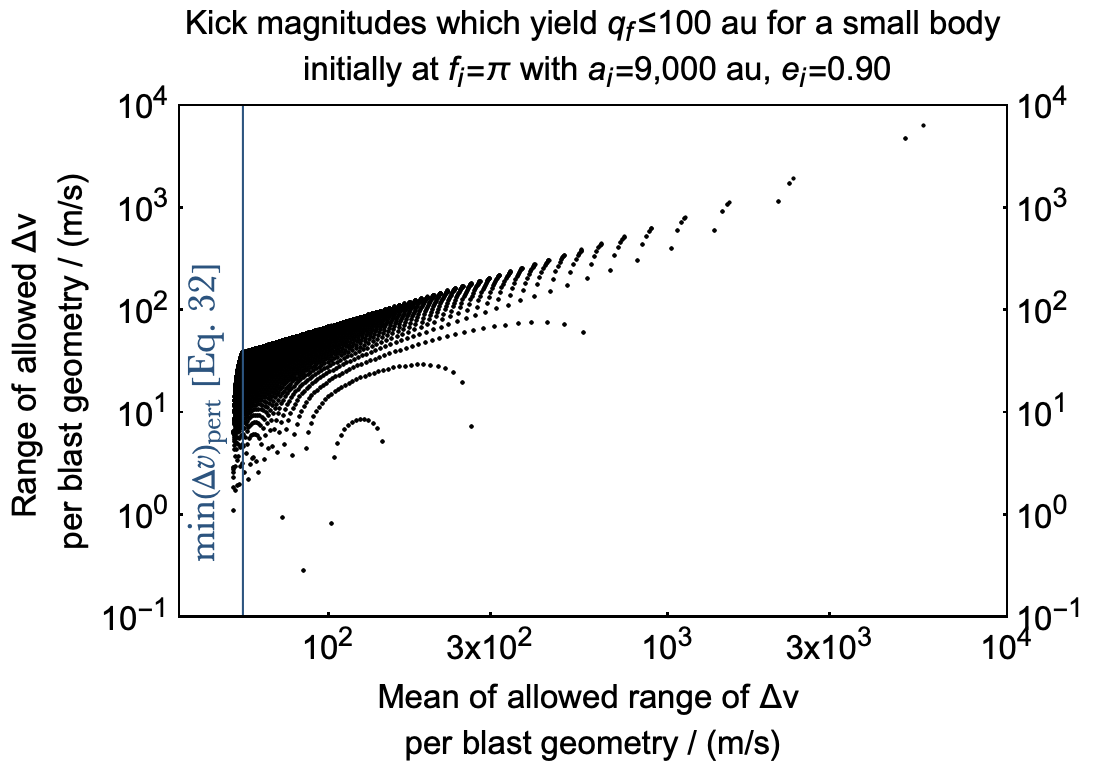}
\includegraphics[width=9.0cm]{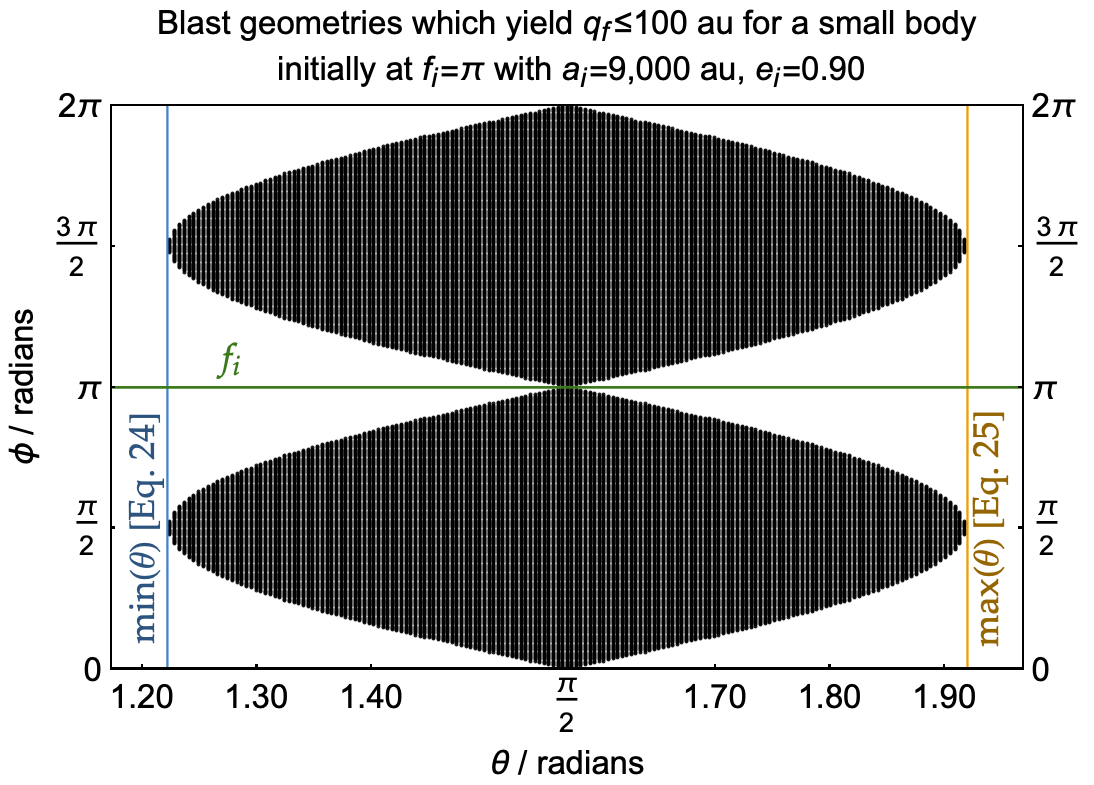}
}
\centerline{}
\centerline{
\includegraphics[width=9.0cm]{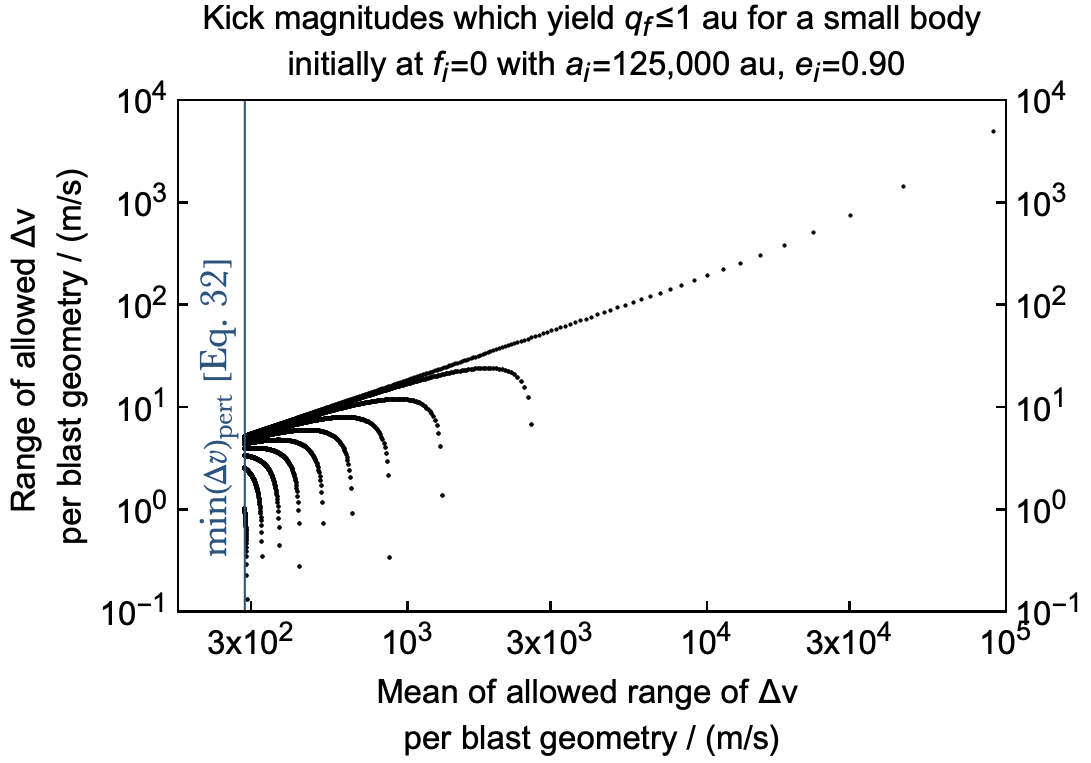}
\includegraphics[width=9.0cm]{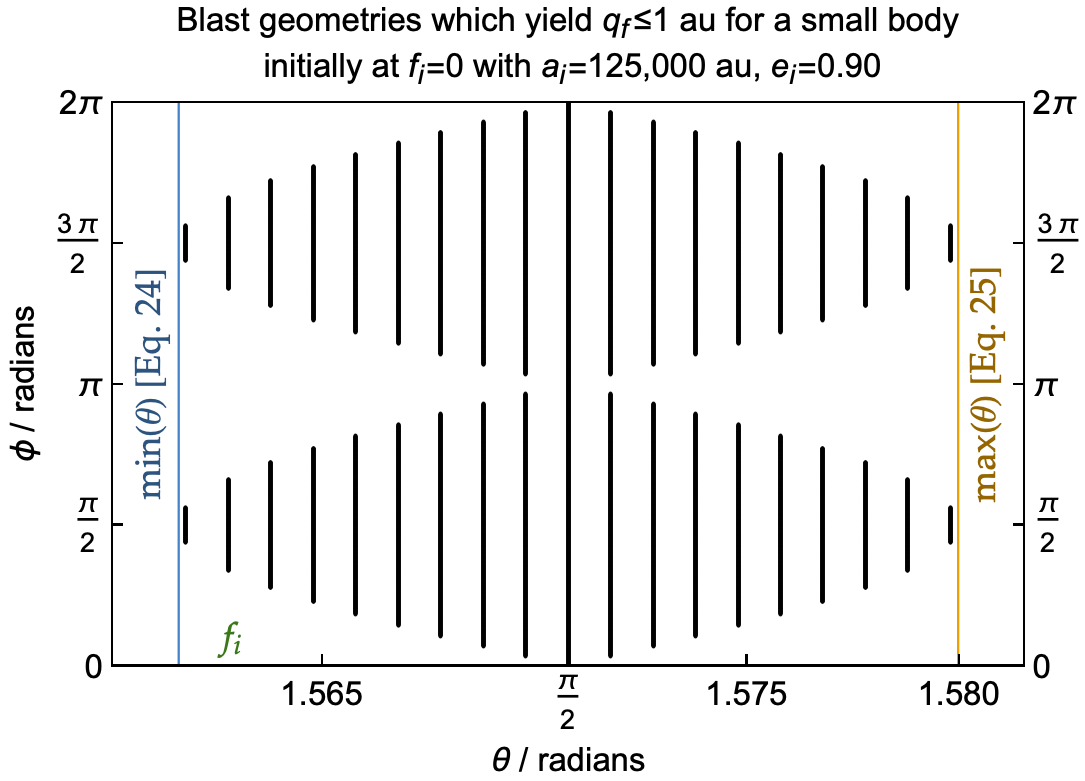}
}
\caption{
Numerical tests to compare with the values of equations (\ref{mintheta}), (\ref{maxtheta}), (\ref{exactP}), (\ref{Ptotal}), and (\ref{minvlimit}). Each row corresponds to a different simulation where ($1.0 \times 10^6$, $2.25 \times 10^5$, $4.0 \times 10^6$) supernova blast wave geometries were sampled, respectively. The right panels display only geometries that achieved a post-blast orbital pericentre $q_{\rm f}$ as is specified in the plot titles. The probabilities resulting from equations (\ref{exactP}) and (\ref{Ptotal}), which are not shown on the plots, were within a few per cent of the numerical results (which are resolution limited) in each case. The middle-left plot (with $q_{\rm i} = 900$~au) demonstrates where the correspondence with equation (\ref{minvlimit}) begins to break down because of the large value of $q_{\rm f}/q_{\rm i}$. 
}
\label{numtestsFig}
\end{figure*}

For all three numerical tests, I present plots demonstrating the agreement of equations (\ref{mintheta}), (\ref{maxtheta}) and (\ref{minvlimit}) with the outputs in Fig. \ref{numtestsFig}. Each point on each plot indicates geometries and kick velocities that succeeded in achieving $q_{\rm f} \le 100$~au or $q_{\rm f} \le 1$~au. The density of points is a reflection of both the number of blast geometries sampled and the system architectures tested. The equations are particularly useful for extreme values of $q_{\rm f}/a_{\rm i}$, which introduce higher errors in the numerical simulations.

%%%%%%%%%%%%%%%% Figure 
%%%%%%%%%%%%%%%% Figure 

%%%%%%%%%%%%%%%% Figure
\begin{figure*}
\includegraphics[width=17.0cm]{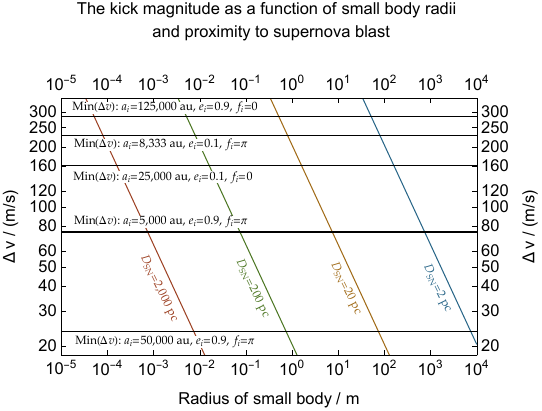}
\caption{
Radii limits of small bodies which are susceptible to inward pericentre thrusts to the perturbation zone as a function of supernova blast wave distance (diagonal lines) and an assumed initial architecture (horizontal lines), from equations (\ref{blastwave}) and (\ref{minvlimit}). The orbital architectures shown in this plot are included in Table \ref{TableNvalFig}.
}
\label{SNdistFig}
\end{figure*}
%%%%%%%%%%%%%%%% Figure

\begin{table*}
\caption {\label{TableNvalFig} User-friendly formulas to compute the number $\mathcal{N}$ (equation \ref{finalN}) of supernova-induced inward pericentre thrusts of small objects with maximum radii of $R_{\rm max}$ (equation \ref{rmaxdep}) to the perturbation zone over the entire cooling age of a white dwarf. The set of values which correspond to the occurrence of repeated pericentre passages (antepenultimate column) are indicated by large red dots on Fig. \ref{KeepBoundFig}. The penultimate column does not make assumptions about the white dwarf cooling age $t_{\rm cool}$ or the time dependence of the supernova rate $\Gamma(t,D_{\rm SN}(t))$ throughout white dwarf cooling, and demonstrates that the supernova blast wave delivery mechanism to old white dwarfs is likely to occur once whenever $\Gamma \sim 0.1-1$~Gyr$^{-1}$ and frequently whenever $\Gamma \gtrsim 1$~Myr$^{-1}$.} 
\begin{tabular}{cccccccccc}
 \multicolumn{6}{c}{{\Large Inputs}} & & \multicolumn{3}{c}{{\Large Outputs}}    \\ 
 \vspace{-0.2cm} \\
 \cline{1-6} \cline{8-10} $D_{\rm SN}$ & $e_i$ & $f_i$ & $r_{\rm esc}/a_i$ & $a_i$ & $r_i$ &  & Repeated pericentre &  $\mathcal{N}$ & $R_{\rm max}$ \\ 
 (pc) &  & (rad) &  & (au) & (au) &  & passages? &  &  \\
 \cline{1-6} \cline{8-10} \\
 2 & 0.9 & $\pi$ & 2.5 &  50,000 & 95,000 & & Yes & $0.046 t_{\rm cool} \Gamma \left(2~{\rm pc} \right)$ & 7.5~km \\
 2 & 0.9 & $\pi$ & 25  &  5,000  & 9,500  & & Yes & $0.15 t_{\rm cool} \Gamma \left(2~{\rm pc} \right)$  & 0.75~km \\ 
 2 & 0.9 & 0     & 1   &  125,000& 12,500 & & No  & $0.029 t_{\rm cool} \Gamma \left(2~{\rm pc} \right)$ & 52~m \\ 
 2 & 0.5 & $\pi$ & 3   &  41,667 & 62,500 & & Yes & $0.025 t_{\rm cool} \Gamma \left(2~{\rm pc} \right)$ & 0.98~km \\
 2 & 0.5 & $\pi$ & 20  &  6,250  & 9,375  & & Yes & $0.065 t_{\rm cool} \Gamma \left(2~{\rm pc} \right)$ & 0.15~km \\
 2 & 0.5 & 0     & 3   &  41,667 & 20,833 & & No  & $0.025 t_{\rm cool} \Gamma \left(2~{\rm pc} \right)$ & 0.11~km \\
 2 & 0.1 & $\pi$ & 15  &  8,333  & 9,177  & & Yes & $0.049 t_{\rm cool} \Gamma \left(2~{\rm pc} \right)$ & 80~m \\
 2 & 0.1 & 0     & 5   &  25,000 & 22,500 & & No  & $0.028 t_{\rm cool} \Gamma \left(2~{\rm pc} \right)$ & 0.16~km \\ 
 \vspace{0.1cm} \\
 20 & 0.9 & $\pi$ & 2.5 &  50,000 & 95,000 & & Yes & $0.046 t_{\rm cool} \Gamma \left(20~{\rm pc} \right)$ & 75~m \\
 20 & 0.9 & $\pi$ & 25  &  5,000  & 9,500  & & Yes & $0.15 t_{\rm cool} \Gamma \left(20~{\rm pc} \right)$  & 7.5~m \\ 
 20 & 0.9 & 0     & 1   &  125,000& 12,500 & & No  & $0.029 t_{\rm cool} \Gamma \left(20~{\rm pc} \right)$ & 0.52~m \\ 
 20 & 0.5 & $\pi$ & 3   &  41,667 & 62,500 & & Yes & $0.025 t_{\rm cool} \Gamma \left(20~{\rm pc} \right)$ & 9.8~m \\
 20 & 0.5 & $\pi$ & 20  &  6,250  & 9,375  & & Yes & $0.065 t_{\rm cool} \Gamma \left(20~{\rm pc} \right)$ & 1.5~m \\
 20 & 0.5 & 0     & 3   &  41,667 & 20,833 & & No  & $0.025 t_{\rm cool} \Gamma \left(20~{\rm pc} \right)$ & 1.1~m \\
 20 & 0.1 & $\pi$ & 15  &  8,333  & 9,177  & & Yes & $0.049 t_{\rm cool} \Gamma \left(20~{\rm pc} \right)$ & 0.80~m \\
 20 & 0.1 & 0     & 5   &  25,000 & 22,500 & & No  & $0.028 t_{\rm cool} \Gamma \left(20~{\rm pc} \right)$ & 1.6~m \\     
 \vspace{0.1cm} \\
 200 & 0.9 & $\pi$ & 2.5 &  50,000 & 95,000 & & Yes & $0.046 t_{\rm cool} \Gamma \left(200~{\rm pc} \right)$ & 0.75~m \\
 200 & 0.9 & $\pi$ & 25  &  5,000  & 9,500  & & Yes & $0.15 t_{\rm cool} \Gamma \left(200~{\rm pc} \right)$  & 75~mm \\ 
 200 & 0.9 & 0     & 1   &  125,000& 12,500 & & No  & $0.029 t_{\rm cool} \Gamma \left(200~{\rm pc} \right)$ & 5.2~mm \\ 
 200 & 0.5 & $\pi$ & 3   &  41,667 & 62,500 & & Yes & $0.025 t_{\rm cool} \Gamma \left(200~{\rm pc} \right)$ & 98~mm \\
 200 & 0.5 & $\pi$ & 20  &  6,250  & 9,375  & & Yes & $0.065 t_{\rm cool} \Gamma \left(200~{\rm pc} \right)$ & 15~mm \\
 200 & 0.5 & 0     & 3   &  41,667 & 20,833 & & No  & $0.025 t_{\rm cool} \Gamma \left(200~{\rm pc} \right)$ & 11~mm \\
 200 & 0.1 & $\pi$ & 15  &  8,333  & 9,177  & & Yes & $0.049 t_{\rm cool} \Gamma \left(200~{\rm pc} \right)$ & 8.0~mm \\
 200 & 0.1 & 0     & 5   &  25,000 & 22,500 & & No  & $0.028 t_{\rm cool} \Gamma \left(200~{\rm pc} \right)$ & 16~mm \\
 \vspace{0.1cm} \\
 2000 & 0.9 & $\pi$ & 2.5 &  50,000 & 95,000 & & Yes & $0.046 t_{\rm cool} \Gamma \left(2~{\rm kpc} \right)$ & 7.5~mm \\
 2000 & 0.9 & $\pi$ & 25  &  5,000  & 9,500  & & Yes & $0.15 t_{\rm cool} \Gamma \left(2~{\rm kpc} \right)$  & 0.75~mm \\ 
 2000 & 0.9 & 0     & 1   &  125,000& 12,500 & & No  & $0.029 t_{\rm cool} \Gamma \left(2~{\rm kpc} \right)$ & 52~$\mu$m \\ 
 2000 & 0.5 & $\pi$ & 3   &  41,667 & 62,500 & & Yes & $0.025 t_{\rm cool} \Gamma \left(2~{\rm kpc} \right)$ & 0.98~mm \\
 2000 & 0.5 & $\pi$ & 20  &  6,250  & 9,375  & & Yes & $0.065 t_{\rm cool} \Gamma \left(2~{\rm kpc} \right)$ & 0.15~mm \\
 2000 & 0.5 & 0     & 3   &  41,667 & 20,833 & & No  & $0.025 t_{\rm cool} \Gamma \left(2~{\rm kpc} \right)$ & 0.11~mm \\
 2000 & 0.1 & $\pi$ & 15  &  8,333  & 9,177  & & Yes & $0.049 t_{\rm cool} \Gamma \left(2~{\rm kpc} \right)$ & 80~$\mu$m \\
 2000 & 0.1 & 0     & 5   &  25,000 & 22,500 & & No  & $0.028 t_{\rm cool} \Gamma \left(2~{\rm kpc} \right)$ & 0.16~mm 
\end{tabular}
\begin{tabbing}
\end{tabbing}
\end{table*}

%%%%%%%%%%%%%%%% Figure
%\begin{figure*}
%\includegraphics[width=13cm]{TableN}
%\caption{
%User-friendly formulas to compute the number $\mathcal{N}$ (equation \ref{finalN}) of supernova-induced inward pericentre thrusts of small objects with maximum radii of $R_{\rm max}$ (equation \ref{rmaxdep}) to the perturbation zone over the entire cooling age of a white dwarf. The third column reflects the values adopted for $r_{\rm imp}$ and $r_{\rm esc}$. The fourth column does not make assumptions about the white dwarf cooling age $t_{\rm cool}$ or the time dependence of the supernova rate $\Gamma(t,D_{\rm SN}(t))$ throughout white dwarf cooling, and demonstrates that the supernova blast wave delivery mechanism to old white dwarfs is likely to occur once whenever $\Gamma \sim 10-100$~Gyr$^{-1}$ and frequently whenever $\Gamma \gtrsim 1$~Myr$^{-1}$.
%}
%\label{TableNvalFig}
%\end{figure*}
%%%%%%%%%%%%%%%% Figure

\section{Relating the kick magnitude to supernova properties}

\subsection{Maximum affected debris size}

Having now bounded $\Delta v$ from below and computed the probability for relevant supernova blast geometries to occur, I can return to equation (\ref{blastwave}) and relate $\Delta v$ to the small body radius $R$ and blast distance $D_{\rm SN}$.

Using the parameter values in the equation as justified in Section 3, I plot the equation as diagonal lines in Fig. \ref{SNdistFig}. Overplotted are horizontal lines corresponding to kick magnitude limits (equation \ref{minvlimit}) from extreme initial architectures. Some of these architectures allow for repeated pericentre passages, and some do not. For those that do, I use values indicated by red dots on Fig. \ref{KeepBoundFig} as approximate limiting cases. The intersection of the horizontal and diagonal lines then bound the maximum debris size for which the debris could eventually be thrust into the perturbation zone. This maximum size is given, with the most important dependencies highlighted, by

\begin{equation}
R_{\rm max}\left(a_{\rm i}, e_{\rm i}, f_{\rm i}, D_{\rm SN} \right) 
\approx 
\frac{3 \gamma E_{\rm SN} a_{\rm i} \left(1 - e_{\rm i}^2\right)}
                         {4 \pi G \rho M_{\star} D_{\rm SN}^2 \left(1 + e_{\rm i}\cos{f_{\rm i}} \right)^2}.
\label{rmaxdep}                         
\end{equation}

\subsection{Frequency of inwards pericentre thrusts}

In order to determine how often small objects which are smaller than $R_{\rm max}$ have their pericentres thrust inwards to the perturbation zone, denote this number of times over a white dwarf cooling age as $\mathcal{N}$. Then, 

\begin{equation}
\mathcal{N} = \mathcal{P}(a_{\rm i}, e_{\rm i}, q_{\rm f}) \Gamma(t, D_{\rm SN}(t)) t_{\rm cool}
,
\label{finalN}
\end{equation}

\noindent{}where $\mathcal{P}$ is given by equation (\ref{exactP}) or (\ref{Ptotal}), $t_{\rm cool}$ is the cooling age of the white dwarf, and $\Gamma$ is the supernova rate. An accurate treatment of $\Gamma$ would require knowing the kinematical history of the white dwarf in the Milky Way, as well as the time and spatial evolution of all Galactic supernovae, so that each value of $D_{\rm SN}$ per blast can be determined. Given the uncertainty in $\Gamma$, I leave it as a variable, and assume that $D_{\rm SN}$ does not vary across the white dwarf cooling age.

I then compute $\mathcal{N}$ and $R_{\rm max}$ for $q_{\rm f}=100$~{\rm au} and a variety of kick limits, including those in Figs. \ref{KeepBoundFig} and \ref{SNdistFig}. I tabulate and present the results in Table \ref{TableNvalFig}. This tabulation allows one to quickly estimate $\mathcal{N}$ for a white dwarf with a given cooling age and an assumed supernova rate $\Gamma$, along with the accompanying applicable small body radius limit ($R_{\rm max}$).

Although every white dwarf's path through the Galaxy is different, even by adopting a generous range of potential $\Gamma$ values, one may glean some overall conclusions from Table \ref{TableNvalFig}. The first is that supernovae which occur on the scale of kpc or hundreds of pc occur frequently enough (\citealt*{quietal2025} give $\Gamma > 10^4$~Gyr$^{-1}$ for $D_{\rm SN}=$1~kpc) that all micron-sized planetary dust and mm-sized sand and pebbles in exo-Oort clouds are subject to ejection or significant elliptical orbit alteration during each supernova blast. As a result, unless the exo-Oort clouds are continuously replenished with these small bodies (from the interstellar medium, the inner planetary system or the break-up of existing larger bodies in the exo-Oort clouds), then the size distribution in these clouds will quickly become top-heavy.

The second conclusion is that rates of $\Gamma$ of a few per Gyr, which is assumed for $D_{\rm SN} = 20$~pc \citep{quietal2025}, are high enough to allow m-sized boulders to be thrust into the perturbation zone at least once, depending on the specific size of $R$. It is this size regime of boulders where the equations predict the greatest variance in outcomes. Only the smallest comets or asteroids ($<10$~km in size) can be thrust to the inner planetary system, and only in the most extreme orbital configurations. Hence, any larger objects in exo-Oort clouds which are delivered to the perturbation zone must be done so through other effects (e.g. stellar and molecular cloud flybys, Galactic tides, natal white dwarf kicks).

\section{Discussion}

This investigation helps place into context the solar system-based numerical simulations of \cite{smietal2024}, and a comparison with their results may be helpful. 

Their simulations modelled the evolution of a population of initially dynamically excited small objects with $0.9<e_{\rm i}<1.0$ and 10,000~au $<a_{\rm i}<$ 20,000~au, and explicitly included $R$, $\rho$ and $D_{\rm SN}$ as inputs. They adopted input ranges of 10$ \mu$m $\le R \le$ 100~m, $0.5$ g~cm$^{-3} < \rho <$ 1.5 g~cm$^{-3}$ and 10~pc $<D_{\rm SN} <$ 100~pc. They also made assumptions about the initial values of the other orbital elements of the small bodies; these elements do not feature in the development presented here.

Although \cite{smietal2024} did not plot $q_{\rm f}$ distributions, their results illustrate a steeply decreasing distribution of $a_{\rm f}$ from $10^3$~au down to $1$~au, and a steeply decreasing distribution of $e_{\rm f}$ from 1.0 to 0.0. Their lowest positive values of $a_{\rm f}$ must have been achieved only in the case of high initial $e_{\rm i}$ (see equation \ref{minminaf}). Further, their results suggest that a large fraction of the small bodies remain in the Oort cloud region post-blast, just with a set of new orbital parameters. In this sense, tracking the population of these objects over a white dwarf cooling age of Gyrs would require an accounting for the successive generations of blasts, as well as the dynamical churning in the region which occurs in-between blasts.

A fundamental difference between this study and that of \cite{smietal2024} is that main-sequence stars have winds that can push small bodies out of the system. Further, main-sequence stars are much brighter than old white dwarfs, with much higher radiative blow-out sizes (see equation \ref{blowout}). 

These differences would be enhanced when considering giant branch stars, which can have both more powerful winds and much greater luminosities than main-sequence stars \citep{mustill2024}. In fact, during these giant branch phases, the radiative break-up of boulders and small asteroids, when subjected to these winds, can provide a source of planetary debris to the eventual white dwarf exo-Oort cloud \citep{veretal2015,veretal2019}. Partly because observational signatures of planetary nebula-based sources of debris around very hot white dwarfs remain ambiguous \citep{suetal2007,maretal2023,estetal2025}, constraining the amount of freshly generated exo-Oort cloud material around newly born white dwarfs remains challenging.

In fact, detailed population analyses of this material would rely on a comparison of sources and sinks, Galactic trajectories and the structure and evolution of exo-Oort clouds. Further, for boulder-sized objects, supernova radiation could have long-term effects on their orbits, spins and thermal evolution through the Yarkovsky effect, YORP effect and sublimation.

Potential replenishment sources for exo-Oort clouds would include the nascent main-sequence exo-debris discs \citep[e.g.][]{marino2022,pearce2024}, giant branch radiative break-up \citep{veretal2014c,versch2020}, supernova-induced sublimation around the white dwarf \citep{belpav2024}, and a regular stream of interstellar interlopers \citep{jewitt2017,trietal2017,doetal2018,jewsel2023}. Unlike the asteroidal interlopers, which will pass right through an exo-Oort cloud (on Myr timescales), smaller objects are more likely to be captured in the gravitational well through a magnetic or plasma-based source of dissipation \citep{athfie2011,zhaetal2021}.

Potential depletion sinks for exo-Oort cloulds would include giant branch winds and stellar flybys \citep{alcetal1986,paralc1998,veretal2014b,stoetal2015,caihey2017,ocoetal2023,pharei2024}, as well as each nearby supernova blast \citep{smietal2024}. Magnetic field interactions, particularly for the smallest debris particles, may -- just as they act as sources -- could also represent sinks \citep{belraf2010}. These sources and sinks would contribute to the structure of the exo-Oort clouds \citep[e.g.][]{higuchi2020,batnes2024,nesetal2025}, which in turn would evolve according to Galactic tides and molecular cloud flybys as the white dwarfs traverse the Galaxy.

A longstanding argument against exo-Oort cloud material representing the primary source of enrichment, or pollution, of white dwarfs is the chemical mismatch between observed volatile-poor debris in most of these systems (see Section 3.2 of \citealt*{xuetal2024} for a substantial reference list) and the assumed composition of exo-Oort cloud debris. However, \cite{zhaetal2021} suggested that the volatiles from exo-Oort cloud debris could be sublimated away (not even accounting for nearby supernovae), an idea which has since become the subject of more detailed study \citep{levetal2023}. Alternatively, \cite{ocoetal2023} indicated that the paucity of volatile-rich observations could be due to planets within the perturbation zone acting as dynamical barriers to direct accretion onto the white dwarf; for eventual enrichment of the white dwarf, the planets need only deliver post-blast debris into a region where white dwarf radiation could take over \citep{veretal2022}. For rarer and deep enough post-blast incursions into the inner planetary system (see Fig. \ref{ProbFig}), debris may be captured directly into the debris disc surrounding the white dwarf \citep{griver2019}.

\section{Summary}

Planetary debris is observed in the photospheres of stars which have already existed as white dwarfs for $\sim$10~Gyr \citep{elmetal2022}. By this time, and after the upheaval which accompanies giant branch evolution,  primordial planetary debris reservoirs within 100~au will have likely already been depleted significantly. 

An alternative potential source of debris around such old stars are exo-Oort clouds, and one external and common phenomena occurring throughout the history of the Milky Way that could deliver cloud-based debris into inner planetary systems is supernovae. Further, the quick speed of a supernova blast wave, combined with long orbital periods of the debris, allows for the impulse formalism to be applied when modelling the outcome of the debris experiencing the blast wave.

Here I have used this formalism to analytically constrain the prospects for supernova blast waves to propel distant debris into the inner reaches of old white dwarf planetary systems. From this exercise, I have derived expressions for:

\begin{enumerate}

\item The range of allowable blast geometries (equations \ref{condtheta}-\ref{eqforT}).

\,

\item The probability for attaining these geometries (equations \ref{exactP} and \ref{Ptotal}).

\,

\item The minimum magnitude of the kick required to propel the debris into the inner reaches of the planetary systems (equation \ref{minvlimit}).

\,

\item The minimum magnitude of the kick required to alter the initial orbit in a way such that it begins leaking (when the apocentre is outside of the gravitational reach of the white dwarf), preventing repeated post-blast pericentre passages (equation \ref{minleak}).

\,

\item The minimum magnitude of the kick required to change the initial elliptical orbit into a hyperbolic one, preventing repeated post-blast pericentre passages (equation \ref{minbreak}).

\,

\item The initial orbital conditions which allow for a post-blast elliptical orbit to remain entirely within the white dwarf's gravitational sphere of influence (equation \ref{pertcond}) and the prospects for doing so (equations \ref{fcrit}, \ref{frangefrac} and \ref{maxFeq}). In this case, repeated post-blast pericentre passages occur.

\,

\item The maximum radii of debris that is subject to this inwards pericentre thrust (equation \ref{rmaxdep}). 

\end{enumerate}

Then, by defining a supernova rate function and white dwarf cooling age, I determined the number of inward pericentre thrusts that an old white dwarf would experience, along with the applicable upper limit on debris size (Table \ref{TableNvalFig}). This calculation reveals that for a recent Galactic evolution model \citep{quietal2025}, micron-sized dust and mm-sized sand and pebbles are ubiquitously and repeatedly either ejected from the system or are thrust inwards by supernovae. Metre-sized boulders, more likely than not, experience the same behaviour at least once, while objects which are a few orders of magnitude larger are affected similarly for only a small fraction of old white dwarfs. Comets larger than about 10~km could only be delivered in this manner by a supernova which occurs within the exo-Oort cloud itself.

\section*{Acknowledgements}

I thank the reviewer for very helpful comments that have improved the manuscript.

\section*{Data Availability}

All data generated in this paper is available upon reasonable request to the author.

\label{lastpage}

\begin{thebibliography}{99}

 
\bibitem[Akiba et al.(2024)]{akietal2024} Akiba, T., McIntyre, S., \& Madigan, A.-M.\ 2024, ApJL, 
%Tidal Disruption of Planetesimals from an Eccentric Debris Disk Following a White Dwarf Natal Kick, 
966, 1, L4. 
%doi:10.3847/2041-8213/ad394c

\bibitem[Alcock et al.(1986)]{alcetal1986} Alcock, C., Fristrom, C.~C., \& Siegelman, R.\ 1986, ApJ, 
%On the Number of Comets around Other Single Stars, 
302, 462. 
%doi:10.1086/164005

\bibitem[Athanassiadou \& Fields(2011)]{athfie2011} Athanassiadou, T. \& Fields, B.~D.\ 2011, New Astronomy, %Penetration of nearby supernova dust in the inner solar system, 
16, 4, 229. 
%doi:10.1016/j.newast.2010.09.007

\bibitem[Bailey \& Fabrycky(2019)]{baifab2019} Bailey, N. \& Fabrycky, D.\ 2019, AJ, 
%Stellar Flybys Interrupting Planet-Planet Scattering Generates Oort Planets, 
158, 2, 94. 
%doi:10.3847/1538-3881/ab2d2a

\bibitem[Batygin \& Nesvorn{\'y}(2024)]{batnes2024} Batygin, K. \& Nesvorn{\'y}, D.\ 2024, Celestial Mechanics and Dynamical Astronomy, 
%Self-gravitational dynamics within the inner Oort cloud, 
136, 3, 24. 
%doi:10.1007/s10569-024-10195-2

\bibitem[Belyaev \& Rafikov(2010)]{belraf2010} Belyaev, M.~A. \& Rafikov, R.~R.\ 2010, ApJ, 
%The Dynamics of Dust Grains in the Outer Solar System, 
723, 2, 1718. 
%doi:10.1088/0004-637X/723/2/1718

\bibitem[Belousov \& Pavlov(2024)]{belpav2024} Belousov, D.~V. \& Pavlov, A.~K.\ 2024, Icarus, 
%Cometary outbursts in the Oort cloud, 
415, 116066. 
%doi:10.1016/j.icarus.2024.116066

\bibitem[Ben{\'\i}tez et al.(2002)]{benetal2002} Ben{\'\i}tez, N., Ma{\'\i}z-Apell{\'a}niz, J., \& Canelles, M.\ 2002, Physical Review Letters, 
%Evidence for Nearby Supernova Explosions, 
88, 8, 081101. 
%doi:10.1103/PhysRevLett.88.081101

\bibitem[Blouin \& Xu(2022)]{bloxu2022} Blouin, S. \& Xu, S.\ 2022, MNRAS, 
%No evidence for a strong decrease of planetesimal accretion in old white dwarfs, 
510, 1, 1059. 
%doi:10.1093/mnras/stab3446

\bibitem[Bonsor et al.(2011)]{bonetal2011} Bonsor, A., Mustill, A.~J., \& Wyatt, M.~C.\ 2011, MNRAS, 
%Dynamical effects of stellar mass-loss on a Kuiper-like belt, 
414, 2, 930. 
%doi:10.1111/j.1365-2966.2011.18524.x

\bibitem[Bonsor \& Wyatt(2012)]{bonwya2012} Bonsor, A. \& Wyatt, M.~C.\ 2012, MNRAS, 
%The scattering of small bodies in planetary systems: constraints on the possible orbits of cometary material,
420, 4, 2990. 
%doi:10.1111/j.1365-2966.2011.20156.x

\bibitem[Bonsor(2024)]{bonsor2024} Bonsor, A.\ 2024, Encyclopedia of Astrophysics, White Dwarf Systems: the Composition of Exoplanets, arXiv:2409.13294. doi:10.48550/arXiv.2409.13294

\bibitem[Bottke et al.(2005)]{botetal2005} Bottke, W.~F., Durda, D.~D., Nesvorn{\'y}, D., Jedicke, R., Morbidelli, A., Vokrouhlick{\'y}, D., Levison, H.~F.\ 2005, Icarus, 
%Linking the collisional history of the main asteroid belt to its dynamical excitation and depletion, 
179, 1, 63. 
%doi:10.1016/j.icarus.2005.05.017

\bibitem[Bottke et al.(2023)]{botetal2023} Bottke, W.~F. et al.\ 2023, PSJ, 
%The Collisional Evolution of the Primordial Kuiper Belt, Its Destabilized Population, and the Trojan Asteroids,
4, 9, 168. 
%doi:10.3847/PSJ/ace7cd

\bibitem[Breitschwerdt et al.(2016)]{breetal2016} Breitschwerdt, D., Feige, J., Schulreich, M.~M., Avillez, M.~A. De., Dettbarn, C., Fuchs, B.\ 2016, Nature, 
%The locations of recent supernovae near the Sun from modelling $^{60}$Fe transport, 
532, 7597, 73. 
%doi:10.1038/nature17424

\bibitem[Brouwers et al.(2022)]{broetal2022} Brouwers, M.~G., Bonsor, A., \& Malamud, U.\ 2022, MNRAS, 
%A road-map to white dwarf pollution: tidal disruption, eccentric grind-down, and dust accretion, 
509, 2, 2404. 
%doi:10.1093/mnras/stab3009

\bibitem[Burns et al.(1979)]{buretal1979} Burns, J.~A., Lamy, P.~L., \& Soter, S.\ 1979, Icarus, 
%Radiation forces on small particles in the solar system, 
40, 1, 1. 
%doi:10.1016/0019-1035(79)90050-2

\bibitem[Caiazzo \& Heyl(2017)]{caihey2017} Caiazzo, I. \& Heyl, J.~S.\ 2017, MNRAS, 
%Polluting white dwarfs with perturbed exo-comets, 
469, 3, 2750. 
%doi:10.1093/mnras/stx1036

\bibitem[Chen et al.(2019)]{cheetal2019} Chen, D.-C. et al.\ 2019, Nature Astronomy, 
%A power-law decay evolution scenario for polluted single white dwarfs, 
3, 69. 
%doi:10.1038/s41550-018-0609-7

\bibitem[Correa-Otto \& Calandra(2019)]{corcal2019} Correa-Otto, J.~A. \& Calandra, M.~F.\ 2019, MNRAS, 
%Stability in the most external region of the Oort Cloud: evolution of the ejected comets, 
490, 2, 2495. 
%doi:10.1093/mnras/stz2671

\bibitem[Cunningham et al.(2024)]{cunetal2024} Cunningham, T., Tremblay, P.-E., \& W. O'Brien, M.\ 2024, MNRAS,
%Initial-final mass relation from white dwarfs within 40 pc, 
527, 2, 3602. 
%doi:10.1093/mnras/stad3275

\bibitem[Debes et al.(2012)]{debetal2012} Debes, J.~H., Walsh, K.~J., \& Stark, C.\ 2012, ApJ, 
%The Link between Planetary Systems, Dusty White Dwarfs, and Metal-polluted White Dwarfs, 
747, 2, 148. 
%doi:10.1088/0004-637X/747/2/148

\bibitem[Do et al.(2018)]{doetal2018} Do, A., Tucker, M.~A., \& Tonry, J.\ 2018, ApJL, 
%Interstellar Interlopers: Number Density and Origin of `Oumuamua-like Objects, 
855, 1, L10. 
%doi:10.3847/2041-8213/aaae67

\bibitem[Dones et al.(2015)]{donetal2015} Dones, L., Brasser, R., Kaib, N., Rickman, H.\ 2015, Space Science Reviews, 
%Origin and Evolution of the Cometary Reservoirs, 
197, 191

\bibitem[Duncan et al.(1987)]{dunetal1987} Duncan, M., Quinn, T., \& Tremaine, S.\ 1987, AJ, 
%The Formation and Extent of the Solar System Comet Cloud, 
94, 1330. 
%doi:10.1086/114571

\bibitem[Elms et al.(2022)]{elmetal2022} Elms, A.~K., Tremblay, P.-E., G{\"a}nsicke, B.~T., Koester, D., Hollands, M.~A., Gentile Fusillo, N.-P., Cunningham, T., Apps, K.\ 2022, MNRAS, 
%Spectral analysis of ultra-cool white dwarfs polluted by planetary debris, 
517, 3, 4557. 
%doi:10.1093/mnras/stac2908

\bibitem[Estrada-Dorado et al.(2025)]{estetal2025} Estrada-Dorado, S., Guerrero, M.~A., Toal{\'a}, J.~A., Maldonado, R.~F., Lora, V., Vasquez-Torres, D.~A., Chu, Y.-H.\ 2025, MNRAS, 
%Accretion onto WD 2226-210, the central star of the Helix Nebula, 
536, 3, 2477. 
%doi:10.1093/mnras/stae2733


\bibitem[Fields et al.(2020)]{fieetal2020} Fields, B.~D. et al.\ 2020, Proceedings of the National Academy of Science, 
%Supernova triggers for end-Devonian extinctions, 
117, 35, 21008. 
%doi:10.1073/pnas.2013774117

\bibitem[Firestone(2014)]{firestone2014} Firestone, R.~B.\ 2014, ApJ, 
%Observation of 23 Supernovae That Exploded <300 pc from Earth during the past 300 kyr, 
789, 1, 29. 
%doi:10.1088/0004-637X/789/1/29

\bibitem[Frewen \& Hansen(2014)]{frehan2014} Frewen, S.~F.~N. \& Hansen, B.~M.~S.\ 2014, MNRAS, 
%Eccentric planets and stellar evolution as a cause of polluted white dwarfs, 
439, 3, 2442. 
%doi:10.1093/mnras/stu097

\bibitem[Gentile Fusillo et al.(2021)]{genetal2021} Gentile Fusillo, N.~P. et al.\ 2021, MNRAS, 
%A catalogue of white dwarfs in Gaia EDR3, 
508, 3, 3877. 
%doi:10.1093/mnras/stab2672

\bibitem[Gladman et al.(1997)]{gladman1997} Gladman, B.~J. et al.\ 1997, Science, 
%Dynamical lifetimes of objects injected into asteroid belt resonances, 
277, 197. 
%doi:10.1126/science.277.5323.197

\bibitem[Granvik et al.(2017)]{graetal2017} Granvik, M., Morbidelli, A., Vokrouhlick{\'y}, D., Bottke, W.~F., Nesvorn\'{y}, D., Jedicke, R.\ 2017, A\&A, 
%Escape of asteroids from the main belt, 
598, A52. 
%doi:10.1051/0004-6361/201629252

\bibitem[Grishin \& Veras(2019)]{griver2019} Grishin, E. \& Veras, D.\ 2019, MNRAS, 
%Embedding planetesimals into white dwarf discs from large distances, 
489, 1, 168. 
%doi:10.1093/mnras/stz2148

\bibitem[Herbst et al.(2021)]{heretal2021} Herbst, W., Greenwood, J.~P., \& Yap, T.~E.\ 2021, PSJ, 
%The Macroporosity of Rubble Pile Asteroid Ryugu and Implications for the Origin of Chondrules, 
2, 3, 110. 
%doi:10.3847/PSJ/abf7c0

\bibitem[Higuchi \& Kokubo(2015)]{higkok2015} Higuchi, A. \& Kokubo, E.\ 2015, AJ, 
%Effect of Stellar Encounters on Comet Cloud Formation, 
150, 1, 26. 
%doi:10.1088/0004-6256/150/1/26

\bibitem[Higuchi(2020)]{higuchi2020} Higuchi, A.\ 2020, AJ, 
%Anisotropy of Long-period Comets Explained by Their Formation Process, 
160, 3, 134. 
%doi:10.3847/1538-3881/aba94d

\bibitem[Hollands et al.(2018)]{holetal2018} Hollands, M.~A., G{\"a}nsicke, B.~T., \& Koester, D.\ 2018, MNRAS, 
%Cool DZ white dwarfs II: compositions and evolution of old remnant planetary systems, 
477, 1, 93. 
%doi:10.1093/mnras/sty592

\bibitem[Jackson et al.(2014)]{jacetal2014} Jackson, A.~P., Wyatt, M.~C., Bonsor, A., Veras, D. \ 2014, MNRAS, 
%Debris froms giant impacts between planetary embryos at large orbital radii, 
440, 4, 3757. 
%doi:10.1093/mnras/stu476

\bibitem[Jewitt et al.(2017)]{jewitt2017} Jewitt, D., Luu, J., Rajagopal, J., Kotulla, R., Ridgway, S., Liu, W., Augusteijn, T.\ 2017, ApJL, 
%Interstellar Interloper 1I/2017 U1: Observations from the NOT and WIYN Telescopes, 
850, 2, L36. 
%doi:10.3847/2041-8213/aa9b2f

\bibitem[Jewitt \& Seligman(2023)]{jewsel2023} Jewitt, D. \& Seligman, D.~Z.\ 2023, ARA\&A, 
%The Interstellar Interlopers, 
61, 197. 
%doi:10.1146/annurev-astro-071221-054221

\bibitem[Kilic et al.(2020)]{kiletal2020} Kilic, M., Bergeron, P., Kosakowski, A., Brown, W.~R., Ag\"{u}eros, M.~A., Blouin, S. \ 2020, ApJ, 
%The 100 pc White Dwarf Sample in the SDSS Footprint, 
898, 1, 84. 
%doi:10.3847/1538-4357/ab9b8d

\bibitem[Levine et al.(2023)]{levetal2023} Levine, W.~G., Taylor, A.~G., Seligman, D.~Z., Hoover, D.~J., Jedicke, R., Bergner, J.~B., Laughlin, G.~P. \ 2023, PSJ, 
%Interstellar Comets from Post-main-sequence Systems as Tracers of Extrasolar Oort Clouds, 
4, 7, 124. 
%doi:10.3847/PSJ/acdf58

\bibitem[Li et al.(2022)]{lietal2022} Li, D., Mustill, A.~J., \& Davies, M.~B.\ 2022, ApJ, 
%Metal Pollution of the Solar White Dwarf by Solar System Small Bodies, 
924, 2, 61. 
%doi:10.3847/1538-4357/ac33a8

\bibitem[Li et al.(2025)]{lietal2025} Li, Y., Bonsor, A., Shorttle, O., Rogers, L.~K.\ 2025, MNRAS, 537, 2, 2214. doi:10.1093/mnras/staf182

\bibitem[Manser et al.(2024)]{manetal2024} Manser, C.~J. et al.\ 2024, MNRAS, 
%The frequency of metal enrichment of cool helium-atmosphere white dwarfs using the DESI early data release, 
531, 1, L27. 
%doi:10.1093/mnrasl/slae026

\bibitem[Marino et al.(2018)]{maretal2018} Marino, S., Bonsor, A., Wyatt, M.~C., Kral, Q.\ 2018, MNRAS, 
%Scattering of exocomets by a planet chain: exozodi levels and the delivery of cometary material to inner planets, 
479, 2, 1651. 
%doi:10.1093/mnras/sty1475

\bibitem[Marino(2022)]{marino2022} Marino, S.\ 2022, book ``Planetary Systems Now", to be published by World Scientific, Edited by: Luisa M Lara and David Jewitt, Planetesimal/Debris discs, Pages:381–408, arXiv:2202.03053. doi:10.48550/arXiv.2202.03053

\bibitem[Marois et al.(2010)]{maretal2010} Marois, C., Zuckerman, B., Konopacky, Q.~M., Macintosh, B., Barman, T.\ 2010, Nature, 
%Images of a fourth planet orbiting HR 8799, 
468, 7327, 1080. 
%doi:10.1038/nature09684

\bibitem[Marshall et al.(2023)]{maretal2023} Marshall, J.~P., Ertel, S., Birtcil, E., Villaver, E., Kemper, F., Boffin, H., Scicluna, P., Kamath, D.\ 2023, AJ, 
%Evidence for the Disruption of a Planetary System During the Formation of the Helix Nebula, 
165, 1, 22. 
%doi:10.3847/1538-3881/ac9d90

\bibitem[Miller \& Fields(2022)]{milfie2022} Miller, J.~A. \& Fields, B.~D.\ 2022, ApJ, 
%Heliospheric Compression Due to Recent Nearby Supernova Explosions, 
934, 1, 32. 
%doi:10.3847/1538-4357/ac77f1

\bibitem[Morbidelli(1997)]{morbidelli1997} Morbidelli, A.\ 1997, Icarus, 
%Chaotic Diffusion and the Origin of Comets from the 2/3 Resonance in the Kuiper Belt, 
127, 1, 1. 
%doi:10.1006/icar.1997.5681

\bibitem[Mustill(2024)]{mustill2024} Mustill, A.\ 2024, Encyclopedia of Astrophysics, Giant branch planetary systems: Dynamical and radiative evolution, arXiv:2405.09399. doi:10.48550/arXiv.2405.09399

\bibitem[Mustill et al.(2018)]{musetal2018} Mustill, A.~J., Villaver, E., Veras, D., G\"{a}nsicke, B.~T., Bonsor, A.\ 2018, MNRAS,
%Unstable low-mass planetary systems as drivers of white dwarf pollution, 
476, 3, 3939. 
%doi:10.1093/mnras/sty446

\bibitem[Nesvorn{\'y} et al.(2025)]{nesetal2025} Nesvorn{\'y}, D., Dones, L., Vokrouhlick{\'y}, D., Levison, H.~F., Beaug{\'e}, C., Faherty, J., Emmart, C., Parker, J.~P.\ 2025, ApJ, 
%A Spiral Structure in the Inner Oort Cloud, 
983, 1, 74. 
%doi:10.3847/1538-4357/adbf9b

\bibitem[O'Brien et al.(2024)]{obretal2024} O'Brien, M.~W. et al.\ 2024, MNRAS,
%The 40 pc sample of white dwarfs from Gaia, 
527, 3, 8687. 
%doi:10.1093/mnras/stad3773

\bibitem[O'Connor et al.(2022)]{ocoetal2022} O'Connor, C.~E., Teyssandier, J., \& Lai, D.\ 2022, MNRAS, 
%Secular chaos in white dwarf planetary systems: origins of metal pollution and short-period planetary companions, 
513, 3, 4178. 
%doi:10.1093/mnras/stac1189

\bibitem[O'Connor et al.(2023)]{ocoetal2023} O'Connor, C.~E., Lai, D., \& Seligman, D.~Z.\ 2023, MNRAS, 
%On the pollution of white dwarfs by exo-Oort cloud comets, 
524, 4, 6181. 
%doi:10.1093/mnras/stad2281

\bibitem[Parriott \& Alcock(1998)]{paralc1998} Parriott, J. \& Alcock, C.\ 1998, ApJ, 
%On the Number of Comets Around White Dwarf Stars: Orbit Survival During the Late Stages of Stellar Evolution, 
501, 1, 357. 
%doi:10.1086/305802

\bibitem[Pearce(2024)]{pearce2024} Pearce, T.~D.\ 2024, Encyclopedia of Astrophysics, Debris disks around main-sequence stars, arXiv:2403.11804. doi:10.48550/arXiv.2403.11804

\bibitem[Pe{\~n}arrubia(2023)]{penarrubia2023} Pe{\~n}arrubia, J.\ 2023, MNRAS, 
%A halo of trapped interstellar matter surrounding the Solar system, 
519, 2, 1955. 
%doi:10.1093/mnras/stac3642

\bibitem[Pham \& Rein(2024)]{pharei2024} Pham, D. \& Rein, H.\ 2024, MNRAS, 
%Polluting white dwarfs with Oort cloud comets, 
530, 3, 2526. 
%doi:10.1093/mnras/stae986

\bibitem[Portegies Zwart et al.(2021)]{poretal2021} Portegies Zwart, S., Torres, S., Cai, M.~X., Brown, A.~G.~A.\ 2021, A\&A, 
%Oort cloud Ecology. II. the chronology of the formation of the Oort cloud, 
652, A144. 
%doi:10.1051/0004-6361/202040096

\bibitem[Quintana et al.(2025)]{quietal2025} Quintana, A.~L., Wright, N.~J., \& Mart{\'\i}nez Garc{\'\i}a, J.\ 2025, MNRAS, 
%A census of OB stars within 1 kpc and the star formation and core collapse supernova rates of the Milky Way,
538, 3, 1367. 
%doi:10.1093/mnras/staf083

\bibitem[Raymond et al.(2023)]{rayetal2023} Raymond, S.~N., Izidoro, A., \& Kaib, N.~A.\ 2023, MNRAS, 
%Oort cloud (exo)planets, 
524, 1, L72. 
%doi:10.1093/mnrasl/slad079

\bibitem[Rebassa-Mansergas et al.(2019)]{rebetal2019} Rebassa-Mansergas, A., Solano, E., Xu, S., Rodrigo, C., Jim\'{e}nez-Esteban, F.~M., Torres, S.\ 2019, MNRAS, 
%Infrared-excess white dwarfs in the Gaia 100 pc sample, 
489, 3, 3990. 
%doi:10.1093/mnras/stz2423

\bibitem[Roberts et al.(2025)]{robetal2025} Roberts, E.~K., Tremblay, P.-E., O'Brien, M.~W., B\'{e}dard, Antoine, Cunningham, T., Byrne, C.~M., Cukanovaite, E.\ 2025, MNRAS, 
%Comparison of methods used to derive the Galactic star formation history from white dwarf samples, 
538, 4, 2548. 
%doi:10.1093/mnras/staf434

\bibitem[Rodet \& Lai(2024)]{rodlai2024} Rodet, L. \& Lai, D.\ 2024, MNRAS, 
%Planet-driven scatterings of planetesimals into a star: probability, time-scale, and applications, 
527, 4, 11664. 
%doi:10.1093/mnras/stad3905

\bibitem[Ruiter \& Seitenzahl(2025)]{ruisei2025} Ruiter A.~J., Seitenzahl I.~R., 2025, A\&ARv, 33, 1. doi:10.1007/s00159-024-00158-9

\bibitem[Saillenfest(2020)]{saillenfest2020} Saillenfest, M.\ 2020, Celestial Mechanics and Dynamical Astronomy, 
%Long-term orbital dynamics of trans-Neptunian objects, 
132, 2, 12. 
%doi:10.1007/s10569-020-9954-9

\bibitem[Shannon et al.(2015)]{shaetal2015} Shannon, A., Jackson, A.~P., Veras, D., Wyatt, M.\ 2015, MNRAS, 
%Eight billion asteroids in the Oort cloud, 
446, 2, 2059. 
%doi:10.1093/mnras/stu2267

\bibitem[Smallwood et al.(2018)]{smaetal2018} Smallwood, J.~L., Martin, R.~G., Livio, M., Lubow, S.~H.\ 2018, MNRAS, 
%White dwarf pollution by asteroids from secular resonances, 
480, 1, 57. 
%doi:10.1093/mnras/sty1819

\bibitem[Smith et al.(2024)]{smietal2024} Smith, L., Miller, J.~A., \& Fields, B.~D.\ 2024, ApJL, 
%Nearby Supernova and Cloud Crossing Effects on the Orbits of Small Bodies in the Solar System, 
974, 2, L29. 
%doi:10.3847/2041-8213/ad7e1a

\bibitem[Stern \& Shull(1988)]{steshu1988} Stern, S.~A. \& Shull, J.~M.\ 1988, Nature, 
%The influence of supernovae and passing stars on comets in the Oort cloud, 
332, 6163, 407. 
%doi:10.1038/332407a0

\bibitem[Stone et al.(2015)]{stoetal2015} Stone, N., Metzger, B.~D., \& Loeb, A.\ 2015, MNRAS, 
%Evaporation and accretion of extrasolar comets following white dwarf kicks, 
448, 1, 188. 
%doi:10.1093/mnras/stu2718

\bibitem[Su et al.(2007)]{suetal2007} Su, K.~Y.~L. et al.\ 2007, ApJL, 
%A Debris Disk around the Central Star of the Helix Nebula?, 
657, 1, L41. 
%doi:10.1086/513018

\bibitem[Tiscareno \& Malhotra(2009)]{tismal2009} Tiscareno, M.~S. \& Malhotra, R.\ 2009, AJ, 
%Chaotic Diffusion of Resonant Kuiper Belt Objects, 
138, 3, 827. 
%doi:10.1088/0004-6256/138/3/827

\bibitem[Torres et al.(2021)]{toretal2021} Torres, S., Rebassa-Mansergas, A., Camisassa, M.~E., Raddi, R.\ 2021, MNRAS, 
%The Gaia DR2 halo white dwarf population: the luminosity function, mass distribution, and its star formation history, 
502, 2, 1753. 
%doi:10.1093/mnras/stab079

\bibitem[Trilling et al.(2017)]{trietal2017} Trilling, D.~E. et al.\ 2017, ApJL, 
%Implications for Planetary System Formation from Interstellar Object 1I/2017 U1 (`Oumuamua), 
850, 2, L38. 
%doi:10.3847/2041-8213/aa9989

\bibitem[Veras et al.(2011)]{veretal2011} Veras, D., Wyatt, M.~C., Mustill, A.~J., Bonsor, A., Eldridge, J.~J.\ 2011, MNRAS, 
%The great escape: how exoplanets and smaller bodies desert dying stars, 
417, 3, 2104. 
%doi:10.1111/j.1365-2966.2011.19393.x

\bibitem[Veras \& Evans(2013)]{vereva2013} Veras, D. \& Evans, N.~W.\ 2013, MNRAS, 
%Exoplanets beyond the Solar neighbourhood: Galactic tidal perturbations, 
430, 1, 403. 
%doi:10.1093/mnras/sts647

\bibitem[Veras et al.(2014a)]{veretal2014a} Veras, D., Evans, N.~W., Wyatt, M.~C., Tout, A.~C.\ 2014a, MNRAS, 
%The great escape - III. Placing post-main-sequence evolution of planetary and binary systems in a Galactic context, 
437, 2, 1127. 
%doi:10.1093/mnras/stt1905

\bibitem[Veras et al.(2014b)]{veretal2014b} Veras, D., Shannon, A., \& G{\"a}nsicke, B.~T.\ 2014b, MNRAS, 
%Hydrogen delivery onto white dwarfs from remnant exo-Oort cloud comets, 
445, 4, 4175. 
%doi:10.1093/mnras/stu2026

\bibitem[Veras et al.(2014c)]{veretal2014c} Veras, D., Jacobson, S.~A., \& G{\"a}nsicke, B.~T.\ 2014c, MNRAS, 
%Post-main-sequence debris from rotation-induced YORP break-up of small bodies, 
445, 3, 2794. 
%doi:10.1093/mnras/stu1926

\bibitem[Veras et al.(2015)]{veretal2015} Veras, D., Eggl, S., \& G{\"a}nsicke, B.~T.\ 2015, MNRAS, 
%The orbital evolution of asteroids, pebbles and planets from giant branch stellar radiation and winds, 
451, 3, 2814. 
%doi:10.1093/mnras/stv1047

\bibitem[Veras et al.(2019)]{veretal2019} Veras, D., Higuchi, A., \& Ida, S.\ 2019, MNRAS, 
%Speeding past planets? Asteroids radiatively propelled by giant branch Yarkovsky effects, 
485, 1, 708. 
%doi:10.1093/mnras/stz421

\bibitem[Veras \& Scheeres(2020)]{versch2020} Veras, D. \& Scheeres, D.~J.\ 2020, MNRAS, 
%Post-main-sequence debris from rotation-induced YORP break-up of small bodies - II. Multiple fissions, internal strengths, and binary production, 
492, 2, 2437. 
%doi:10.1093/mnras/stz3565

\bibitem[Veras \& Hinkley(2021)]{verhin2021} Veras, D. \& Hinkley, S.\ 2021, MNRAS, 
%The post-main-sequence fate of the HR 8799 planetary system, 
505, 2, 1557. 
%doi:10.1093/mnras/stab1311

\bibitem[Veras et al.(2022)]{veretal2022} Veras, D., Birader, Y., \& Zaman, U.\ 2022, MNRAS, 
%Orbit decay of 2-100 au planetary remnants around white dwarfs with no gravitational assistance from planets, 
510, 3, 3379. 
%doi:10.1093/mnras/stab3490

\bibitem[Veras \& Rosengren(2023)]{verros2023} Veras, D. \& Rosengren, A.~J.\ 2023, MNRAS, 
%The smallest planetary drivers of white dwarf pollution, 
519, 4, 6257. 
%doi:10.1093/mnras/stad130

\bibitem[Veras et al.(2024)]{veretal2024} Veras, D., Mustill, A.~J., \& Bonsor, A.\ 2024, Reviews in Mineralogy and Geochemistry, The Evolution and Delivery of Rocky Extra-Solar Materials to White Dwarfs, Volume 90 on "Exoplanets: Compositions, Mineralogy, and Evolution" edited by Natalie Hinkel, Keith Putirka, and Siyi Xu; doi:10.2138/rmg.2024.90.05

\bibitem[Wajer et al.(2024)]{wajetal2024} Wajer, P., Rickman, H., Kowalski, B., Wi{\'s}niowski, T.\ 2024, Icarus, 
%Oort Cloud and sednoid formation in an embedded cluster, I: Populations and size distributions, 
415, 116065. 
%doi:10.1016/j.icarus.2024.116065

\bibitem[Wallner et al.(2016)]{waletal2016} Wallner, A. et al.\ 2016, Nature, 
%Recent near-Earth supernovae probed by global deposition of interstellar radioactive $^{60}$Fe, 
532, 7597, 69. 
%doi:10.1038/nature17196

\bibitem[Weissman \& Levison(1997)]{weilev1997} Weissman, P.~R. \& Levison, H.~F.\ 1997, ApJL, 
%Origin and Evolution of the Unusual Object 1996 PW: Asteroids from the Oort Cloud?, 
488, 2, L133. 
%doi:10.1086/310940

\bibitem[Williams et al.(2024)]{wiletal2024} Williams, J.~T., G{\"a}nsicke, B.~T., Swan, A., O'Brien, M.~W., Izquierdo, P., Cutolo, A.-M., Cunningham, T.\ 2024, A\&A, 
%PEWDD: A database of white dwarfs enriched by exo-planetary material, 
691, A352. 
%doi:10.1051/0004-6361/202450509

\bibitem[Wyatt et al.(2017)]{wyaetal2017} Wyatt, M.~C., Bonsor, A., Jackson, A.~P., Shannon, A.\ 2017, MNRAS, 
%How to design a planetary system for different scattering outcomes: giant impact sweet spot, maximizing exocomets, scattered discs, 
464, 3, 3385. 
%doi:10.1093/mnras/stw2633

\bibitem[Xu et al.(2024)]{xuetal2024} Xu, S., Rogers, L.~K., \& Blouin, S.\ 2024, Reviews in Mineralogy and Geochemistry, The chemistry of extra-solar materials from white dwarf planetary systems, Volume 90 on "Exoplanets: Compositions, Mineralogy, and Evolution" edited by Natalie Hinkel, Keith Putirka, and Siyi Xu; arXiv:2404.15425. doi:10.48550/arXiv.2404.15425

\bibitem[Zhang et al.(2021)]{zhaetal2021} Zhang, Y., Liu, S.-F., \& Lin, D.~N.~C.\ 2021, ApJ, 
%Orbital Migration and Circularization of Tidal Debris by Alfv{\'e}n-wave Drag: Circumstellar Debris and Pollution around White Dwarfs, 
915, 2, 91. 
%doi:10.3847/1538-4357/ac00ae


\end{thebibliography}
\end{document}